\documentclass[twocolumn,superscriptaddress]{revtex4}
\usepackage{bm}
\usepackage{amsmath, amsfonts, amssymb, graphicx, color, ulem, cancel}
\graphicspath{{../}}
\renewcommand{\thetable}{\arabic{table}}

\renewcommand{\eqref}[1]{Eq.~(\ref{#1})}

\newcommand{\beq}{\begin{equation}}
\newcommand{\eeq}{\end{equation}}
\newcommand{\beqs}{\begin{equation*}}
\newcommand{\eeqs}{\end{equation*}}
\newcommand{\beqn}{\begin{eqnarray}}
\newcommand{\eeqn}{\end{eqnarray}}
\newcommand{\beqns}{\begin{eqnarray*}}
\newcommand{\eeqns}{\end{eqnarray*}}

\newcommand{\RV}{}

\newcommand{\sifigmicelf}{SI Fig.~S1}
\newcommand{\sifigsme}{SI Fig.~S2}

\newcommand{\sifigeta}{SI Fig.~S4}
\newcommand{\sifigsingleisingmh}{SI Fig.~S5}
\newcommand{\sifigemparams}{SI Fig.~S6}
\newcommand{\sifigemks}{SI Fig.~S7}
\newcommand{\sifigemvst}{SI Fig.~S8}

\newcommand{\sifigemvsks}{SI Fig.~S10}
\newcommand{\sifigggdmice}{SI Fig.~S11}
\newcommand{\sifigglmsmeparam}{SI Fig.~S12}
\newcommand{\sifigggdmiceautocorr}{SI Fig.~S13}
\newcommand{\sitable}{SI Table.~S1}

\newcommand{\sisecggdsingleising}{SI Section 1B}
\newcommand{\sisecggdmultiising}{SI Section 1C}
\newcommand{\sisecggdsinglepotts}{SI Section 1D}

\newcommand{\sisecemheu}{SI Section 3D}

\newcommand{\lastequal}{Corresponding authors. These authors contributed equally.}

\begin{document}

\newcommand{\deftitle}{{Generalized Glauber dynamics for inference in biology}}

\title{\deftitle}

\author{Xiaowen Chen}
\affiliation{Laboratoire de physique de l'\'Ecole normale sup\'erieure,
  CNRS, PSL University, Sorbonne Universit\'e, and Universit\'e 
  Paris Cit\'e, 75005 Paris, France}
\author{Maciej Winiarski}
\affiliation{Nencki-EMBL Partnership for Neural Plasticity and Brain Disorders - BRAINCITY, Nencki Institute of Experimental Biology of Polish Academy of Sciences, Pasteur 3 Street, 02-093 Warsaw, Poland}
\author{Alicja Pu{\'s}cian}
\affiliation{Nencki-EMBL Partnership for Neural Plasticity and Brain Disorders - BRAINCITY, Nencki Institute of Experimental Biology of Polish Academy of Sciences, Pasteur 3 Street, 02-093 Warsaw, Poland}
\author{Ewelina Knapska}
\affiliation{Nencki-EMBL Partnership for Neural Plasticity and Brain Disorders - BRAINCITY, Nencki Institute of Experimental Biology of Polish Academy of Sciences, Pasteur 3 Street, 02-093 Warsaw, Poland}
\author{Aleksandra M. Walczak}
\thanks{\lastequal}
\affiliation{Laboratoire de physique de l'\'Ecole normale sup\'erieure,
  CNRS, PSL University, Sorbonne Universit\'e, and Universit\'e 
  Paris Cit\'e, 75005 Paris, France}
\author{Thierry Mora}
\thanks{\lastequal}
\affiliation{Laboratoire de physique de l'\'Ecole normale sup\'erieure,
  CNRS, PSL University, Sorbonne Universit\'e, and Universit\'e 
  Paris Cit\'e, 75005 Paris, France}

\date{\today}

\begin{abstract}
Large interacting systems in biology often exhibit emergent dynamics, such as coexistence of multiple time scales, manifested by fat tails in the distribution of waiting times. While existing tools in statistical inference, such as maximum entropy models, reproduce the empirical steady state distributions, it remains challenging to learn dynamical models. We present a novel inference method, called {\it generalized Glauber dynamics}. Constructed through a non-Markovian fluctuation dissipation theorem, generalized Glauber dynamics tunes the dynamics of an interacting system, while keeping the steady state distribution fixed. We motivate the need for the method on real data from Eco-HAB, an automated habitat for testing behavior in groups of mice under semi-naturalistic conditions, and present it on simple Ising spin systems. We show its applicability for experimental data, by inferring dynamical models of social interactions in a group of mice that reproduce both its collective behavior and the long tails observed in individual dynamics.

\end{abstract}

\maketitle

\section{Introduction}
From collective information encoding in neurons~\cite{population_eye, gt14plos, population_coding_review, puscian2020nmdar} to emergence dynamics in collective animal motion~\cite{vicsek2012collective, rome2008,couzin2009collective, ouellette2013} and population dynamics in ecological communities~\cite{bacteria_sanchez_gore}, 
 collective behavior emerges from dynamical interaction among individual components. In recent years, large scale data acquisition in precisely controlled experiments~\cite{rome2008b, ecohab, peleg_firefly, worm_whole_brain, neuropixel_mice} allow researchers to address these questions, by constructing statistical and dynamical data-based models that reproduce the correlated activity in spiking and non-spiking neurons~\cite{schneidman06, cocco09retina, ferrari2017random, chen_worm}
and collective animal motion~\cite{bialek2012, higher_order_mice}. 

Living systems are intrinsically dynamic and out of equilibrium, as manifested by the co-existence of multiple timescales and the breaking of Markovian rule in animal behavior~\cite{fractal_fly, greg_worm_lts, vasyl_fly, berman2018measuring} and neuron activities~\cite{marques2020internal, kaplan2020nested}. 
However most recent work
 focuses on inferring the static properties of collective behavior: analyzing the joint probability distributions of the interacting components and relating these global states to functional behavior. In many cases,
inferred pairwise interaction models
succesfully reproduce the correlation structure of the data~\cite{bialek2007rediscovering, roudi2009pairwise, stein2015inferring}, leading to identifying the empirical rules for collective neuronal encoding~\cite{schneidman06, gt14plos, leenoy, maxent_cortex, ohiorhenuan2010sparse}, the interaction structure of bird flocks~\cite{bialek2012, attanasi}, or contact maps of proteins~\cite{weigt08}.
While these approaches do not directly address the dynamics, attempts have been made to reconcile them with dynamical models of neurons \cite{chen_worm, wolf2022emergence} and flocks \cite{mora2016} by using classical rules of equilibrium dynamics.

Recent methods to learn the dynamics of interacting systems focus on extensions to second order dynamics for continuous systems~\cite{frishman_ronceray, fede, antonio, thiede2019galerkin}.  For discrete 
spiking neurons, dynamical inference focuses on reproducing pairwise correlation functions between different timepoints~\cite{marre2009prediction, vasquez2012gibbs, cavagna2014dynamical, mora2015dynamical}, or inferring transition probabilities or causal dependencies~\cite{pillow2008spatio}.  The extension of maximum entropy to reproduce time-delayed cross-correlations (called maximum caliber~\cite{presse2013principles}) is computationally expensive and requires a lot of data to train~\cite{marre2009prediction, vasquez2012gibbs}. 
However,many possible dynamical models can generate the same steady state distribution. More fundamentally, in general the dynamics cannot be automatically related to the steady state distribution, especially if the learned models are out-of-equilibrium. In transition models, the transition rate of a component is given by the history of itself and other inputs, stimuli and
interacting components in the network. {\RV{By incorporating autoregressively generated noise in the transition rate, recent developments in Generalized Master Equation (GME) models have introduced the capacity to encode latent variables~\cite{vroylandt2022likelihood}. }} 
 The \textit{generalized linear model} (GLM), a representative of this class used in neuroscience, successfully reproduces different types of neuronal dynamics and firing patterns in many brain regions~\cite{pillow2008spatio}. However, since GLMs are constructed by definition to predict only the next time point forward, they often generate unstable trajectories and produce inconsistent steady state distributions with respect to the training data~\cite{park2013spectral, hocker2017multistep, gerhard2017stability, mahuas2020new}. 

We propose an inference method, called {\it generalized Glauber dynamics} (GGD), that combines the power of steady state inference with dynamical inference. Constructed through a non-Markovian fluctuation dissipation theorem, the generalized Glauber dynamics tunes the dynamics of an interacting system, while keeping the steady state equilibrium distribution fixed. In practice, this method allows for the inference to be separated into two parts: first, inference of the steady state distribution using maximum entropy models, and then, tuning the dynamics to match the data. The basic idea behind the GGD is similar to generalized Langevin dynamics: coupled degrees of freedom are integrated out to generate an effective memory kernel, such that the dynamics of the system depends on its history. Interestingly, the functional form of the GGD is similar to the GLM but differs dramatically in its link to the steady-state distribution. We demonstrate the power of GGD to predict the  co-localization pattern of groups of socially interacting mice.

\section{Results}

\subsection{Collective behavior of social mice}
\begin{figure*}
\centering
\includegraphics[scale=1]{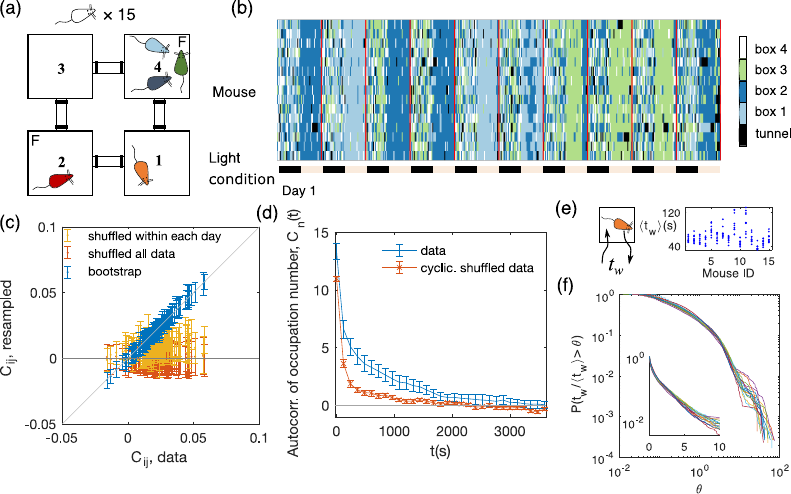}
{\caption{Collective dynamics among social mice.
    (a) The Eco-HAB experimental setup (top view) with C57BL6/J male mice ($N = 15$) placed in four inter-connected chambers. The two chambers with food are labeled by letter \textit{F}. The location of each mouse is recorded using mouse-embedded RFID and antennas at the edge of the tubes (indicated by black bars). (b) Example trace for the co-localization patterns of a group of 15 mice over 10 full days, consisting of alternating dark and light cycles (12 hours). The \textit{red} vertical lines indicate the beginning of each dark cycle. (c) Pairwise connected correlation function of mice colocalization, $C_{ij}$ (with error bar computed by bootstrap). Cyclically shuffled data shows no correlation, while data shuffled within the same day shows a strongly reduced correlation.  (d) Auto-correlation function for the number of mice in a given box, $C_n(t)$, as a function of time difference, computed from the mice position between 13:00 and 19:00 each day, a period of the intensified activity chosen for the presented analysis.  Error bars are the standard error from the mean across 10 days. (e) Mean waiting time for each mouse, defined to be the time a mouse spent staying in a given box before exits. Each dot indicates a different day. (f) Distribution of waiting times normalized by their mean for each mice. Distributions collapse across all mice, and decay slower than exponentially, indicating the existence of long time scales.}\label{fig_micedata}}
\end{figure*}

We studied the interaction structure of groups of animals in a controlled environment.
We analyzed data generated from the Eco-HAB experiment~\cite{ecohab} and presented in~\cite{winiarski2021social}, where a group of $N=15$ freely-moving male mice live in an artificial ecological environment resembling natural burrows (Fig.~\ref{fig_micedata}(a)). 
The Eco-HAB consists of four chambers, two of them with food, connected by tunnels. The experiment lasts for 10 days, with alternating light conditions of darkness and brightness, each lasting 12 hours, to simulate the day-night cycle. Mice are able to behave and interact freely, without any experimental constraints or manipulation.  Each mouse is equipped with a unique radio frequency identification transmitting chips (RFID) that allow for the detection of their position everytime they pass by the 8 recording antennas (marked in black in Fig.~\ref{fig_micedata}(a)). 
The data consists of time points with 1-ms resolution at which each mouse passed a given recording device, which allows us to identify the location of each mouse as a function of time, $\sigma_i(t)=1,2,3$, or $4$ (Fig.~\ref{fig_micedata}(b)).

Mice are nocturnal animals with increased activity during dark periods. For the purpose of method development we focus on dark periods and restrict our analysis to a 6 hour period of stable activity, which consists of the first and bigger of the two nocturnal activity peaks characterizing the used strain of mice, C57BL6/J (see {\sifigmicelf}(a)). 
The individual dynamics are characterized by a basal activity rate of moving between boxes, and by the tendency to explore the next box rather than come back to the previous one, which we term ``roaming'' (see {\sifigmicelf}(b)). 
Dominant mice tend to chase others more frequently~\cite{winiarski2021social}, resulting in chaser-chased dynamics that are significant within a short time scale of a few seconds (see {\sifigmicelf}(c)).

To measure collective behavior at the group scale, we examine the excess frequency of finding two mice in the same box. The mean frequency of mouse $i$ in box $\alpha$ reads $\langle \delta_{\sigma_i, \alpha} \rangle$, where $\delta_{a,b}=1$ if $a=b$, and 0 otherwise. The excess pairwise frequency is given by the correlated pairwise correlation $C_{ij}= \langle \delta_{\sigma_i, \sigma_j} \rangle- \sum_{\alpha} \langle \delta_{\sigma_i, \alpha} \rangle \langle \delta_{\sigma_j, \alpha} \rangle$, where the first average is over all box combinations (Fig.~\ref{fig_micedata}(c)). Non-zero correlations between mice strongly decrease when the data are shuffled across time within the same day, and completely go away when the data are shuffled across days. This suggests that interactions between mice drive the correlation, rather than the environment (i.e. the day).

To study the dynamics of that collective behavior, one can look at the temporal auto-correlation of the total number of mice in each box, $n_{\alpha}(t)$, defined as $C_n(t)={\sum_{\alpha} \langle n_{\alpha}(0) n_{\alpha}(t)\rangle-\langle n_{\alpha} \rangle^2}$.
The excess of this auto-correlation over its counterpart in the shuffled dataset (Fig.~\ref{fig_micedata}(d)) is a signature of group behavior, which slows down the overall dynamics of occupancy by keeping individuals in heavily-occupied boxes longer. This is confirmed by the long tails in the distribution of the waiting time, i.e. the duration between transitioning events (Fig.~\ref{fig_micedata}(e) and (f)).

\subsection{Modeling the steady-state distribution}
These experimental observations suggest collective effects driven by direct interactions between mice that lead to effective phenomena scanning a broad range of timescales.
Our goal is to find a set of effective equations that describe the evolution of the system and are consistent with the properties of the observables  in Fig~\ref{fig_micedata}.
Predicting the full dynamical collective behavior requires defining both the static distribution of box occupancies, and the type of dynamics that governs the transitioning of the mice between boxes. 
We separate the inference problem into two steps: first, we infer the steady-state distribution, $P_\text{\rm s}(\sigma)$ for the macrostates $\sigma=(\sigma_1,\ldots,\sigma_{N})$ ($N=15$), using a maximum entropy approach, and then infer the dynamics while keeping the steady state distribution fixed.

The maximum entropy approach has been applied in a wide range of biological contexts~\cite{schneidman06, cocco09retina, gt14plos,weigt08,antibody10,bialek2012}. It generates approximations to the steady state distribution that match the expectation values of a chosen set of observables while keeping the model otherwise as random as possible.
Here we constrain the co-localization probabilities of all pairs of mice, $C_{ij}$, as well as the single-mice box occupancy functions $\langle\delta_{\sigma_i,\alpha}\rangle$.
The maximum entropy distribution then takes the form of a Boltzmann's law:
\begin{equation}
P_\text{s}(\sigma) = \frac{e^{-E(\sigma)}}{Z_{\rm s}},\ 
  E(\sigma)=-\sum_{i,\alpha} \left(
 h_{i,\alpha} - \sum_{j\neq i} J_{ij} \delta_{\sigma_j, \alpha} 
\right) \delta_{\sigma_i, \alpha},
\label{eq:ss}
\end{equation}
where $h_{i,\alpha} $ and $J_{ij}$ are Lagrange multipliers that must be tuned to satisfy the constraints, $E(\sigma)$ is interpreted as an energy by analogy with statistical mechanics, and $Z_{\rm s}$ enforces normalization.  We fit this model to the Eco-HAB and show that it correctly predicts collective statistics of occupancy that were not fit in the model, such as triad correlations and the probability of pairs of mice to be in a specific box (\sifigsme).

\subsection{Glauber dynamics fails to capture the long-time behavior}~\label{sec_micedynamics}
The same steady-state distribution, \eqref{eq:ss}, can be generated by many different dynamical models. The simplest assumption inspired by statistical physics is to assume that the transition rate of a mouse from one box $\alpha$ to an adjacent one $\beta$ (assuming that transition is instantaneous so that two mice never transition at the same time) is a function of the difference in energies between the ending and starting states, $\Delta E_{i,\alpha\to\beta}(\sigma)$.
Writing the transition rate between adjacent boxes as $W_{i, \alpha \rightarrow \beta} = \mu_i f(\Delta E_{i,\alpha\to\beta})$, where $\mu_i$ is the overall activity of mouse $i$ and $f(\Delta E)$ is a function, a sufficient and necessary condition on $f$ for these rates to admit \eqref{eq:ss} as steady state is given by detailed balance: $f(\Delta E)/f(-\Delta E)=e^{-\Delta E}$.

\begin{figure}
\includegraphics[scale=1]{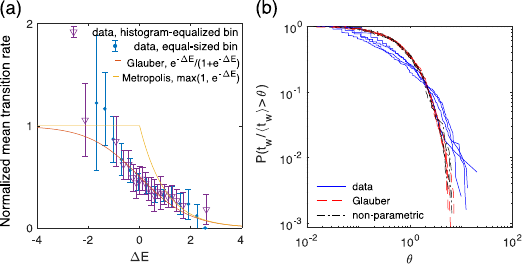}
\caption{Glauber dynamics based on the inferred steady state distribution, \eqref{eq:ss}, fail{\RV{s}} to predict the long-time dynamics of mice. (a) The normalized mean transition rate is well reproduced by Glauber dynamics with the parameters of the steady state model, but not by Metropolis-Hasting dynamics. (b) However, Glauber dynamics with the parameters of the steady state model do not reproduce the long tails of the waiting time distribution. {\RV{The dynamics are also not reproduced by a non-parametric estimation of the transition rates (see SI Fig.~\ref{fig:si:nonparam_transprob}).}} }\label{fig_transition}
\end{figure}

We tested some classical forms for $f(\Delta E)$ on the data (Fig.~\ref{fig_transition}(a)). We found that the empirical normalized transition rates $W_{i, \alpha \rightarrow \beta} / \mu_i$ are
well reproduced by the form of the Glauber dynamics, $f(\Delta E)=1/(1+e^{\Delta E})$, but not by the Metropolis-Hasting prescription, $f(\Delta E)=\min(1,e^{-\Delta E})$.
However, even the Glauber dynamics do not reproduce the long tails in the waiting time distribution, {\RV{nor does a nonparametric form of transition dynamics $f(\Delta E)$ directly estimated from the data (Fig.~\ref{fig_transition}(b), SI Fig.~\ref{fig:si:nonparam_transprob})}}. It did not have to be the case:
while these dynamics are Markovian and memory-less at the group level, long time scales may nonetheless emerge from interactions, as for example during critical slowing down. The failure of the model suggests that, by themselves, these concurrent and co-localized pairwise interactions are not strong enough for such long time scales to emerge. {\RV{ Additionally, the transition probability conditioned on the elapsed time after the last transition exhibits tails (\sifigeta). }}
Together, these results suggest that the dynamics may have a Glauber form, but that additional memory effects must be incorporated.
Here, the term ``memory'' describes how the dynamics depends on the past location time series. 

\subsection{Generalized Glauber dynamics}
Our goal is to add long-term memory effects to the equilibrium
dynamics described above, while keeping maximum entropy distribution valid.
For concreteness, we start from Glauber dynamics, and call the
method the generalized Glauber dynamics (GGD), although the approach
can be generalized to other equilibrium dynamics.

Our approach is general and applies to any group of $N$ correlated, categorical
variables taking $q$ possible values, $\sigma_i=1,\ldots,q$ ($q=4$ in the Eco-HAB).
For simplicity of exposition, here we outline the derivation for a single
binary (Ising) spin ($N = 1$, $q = 2$), $\sigma=\pm 1$ in spin convention. Relating to the Eco-HAB, this
is equivalent to a single mouse placed in an experiment apparatus with two
connected chambers (denoted -1 and +1). The general derivation for arbitrary $N$ and $q$ is given in the SI.

The maximum entropy distribution is given by the energy, $E(\sigma) =
-h \sigma$,
and the Glauber dynamics is defined by the transition rates:
$W(\sigma \rightarrow -\sigma) 
= \mu {e^{-h\sigma}}/
{(e^{h} + e^{-h})}.$
To include memory, we take inspiration from multi-dimensional Markov systems with equilibrium dynamics, such as hidden Markov models and generalized Langevin equations.  
The idea is to consider a larger equilibrium system coupling both the
observed spins and some hidden degrees of freedom. While this augmented
system is Markovian, the subsystem formed by the observed spins
may exhibit memory.

In practice, we couple the spins to a heat bath of harmonic
oscillators (see Fig.~\ref{fig:toy_model_works}(a) for schematics, Ref.~\cite{zwanzig_book} for the standard derivation for
continuous variables and SI for a detailed derivation
for categorical variables). After integrating out the hidden degrees
of freedom,
the transition rates of the GGD take a Glauber-like form, $W(\sigma
\rightarrow -\sigma) = \mu {e^{-h_\text{eff}\sigma}}/{(e^{h_\text{eff}} +
e^{-h_\text{eff}})}$, but with an effective, time-dependent field
\beq\label{eq:heff_single_ising}
h_\text{eff}(t) =  h 
+ \Gamma(0) \sigma(t) 
- \int_0^t dt' \Gamma (t-t') \frac{d \sigma(t')}{dt'} + \xi(t),
\eeq
where the noise correlation satisfies the generalized fluctuation-dissipation
relation 
\beq\label{eq:fdt_single_ising}
\langle \xi(t) \xi(t') \rangle = \Gamma(t-t').
\eeq
$\Gamma(t)$ is an arbitrary function that specifies how the spectrum of
oscillators couples to the spin (see \sisecggdsingleising). 
The second and the third terms depend on the memory kernel, and is defined to be $h_\text{mem}(t)$.

The first term of \eqref{eq:heff_single_ising} is the local
field learned from the maximum entropy model, already present in
classical Glauber dynamics. This terms generalizes readily to the case
of multiple interacting spins as the local field $h_i(\sigma)$ acting on spin
$i$ (defined as half the energy difference between the configurations
with $\sigma_i=-1$ and $\sigma_i=+1$, the other spins being fixed).
The second and the third terms depend on
the history of the spin $\sigma(t)$, and add memory to
the dynamics. The last term the colored noise that results from the
coupling with the memory kernel. When extending to $q$ states, the
fields $h$ and $h_{\rm eff}$ becomes vectors of length $q$, and
$\Gamma(t)$ takes the form of a $q\times q$ matrix coupling the different
states together. Memory kernels $G_{ij}(t)$ may also be added to couple 
different degree of freedom $i$ to the memory of $j$
(see \sisecggdmultiising). 
By construction, the process is reversible, as a subsystem of a
larger equilibrium system including both spins and the oscillator
bath, and its steady state is still given by Boltzmann's law, \eqref{eq:ss}.
The choice for the memory kernel $\Gamma(t)$ is general and can be
chosen from a large family of functions.

\subsection{Range of possible dynamics of GGD}

\begin{figure}[ht!]
\centering
\includegraphics[width=\columnwidth]{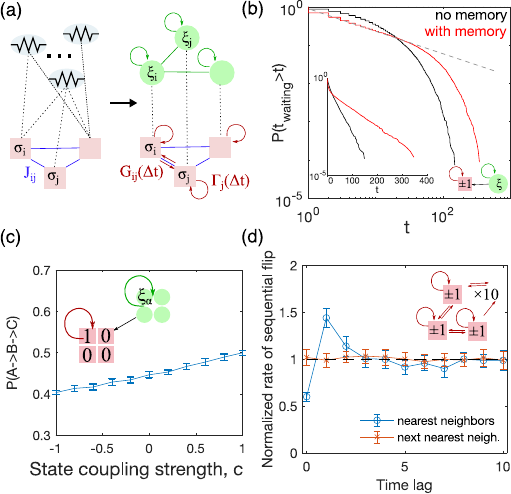}
\caption{Toy models with generalized Glauber dynamics. (a) Schematics
  of the GGD. In addition to classical couplings $J_{ij}$, by considering coupling to an oscillator bath, the
  observed degrees of freedom are coupled to their own memory (through
  kernel $\Gamma_i(t)$), to that of their neighbours (through kernels
  $G_{ij}(t)$), as well as to noise sources $\xi_i(t)$ correlated across time and
  variables. 
  (b)
  shows for a single Ising spin that adding an exponentially decaying memory kernel can create
  long tails in the waiting distribution. (c) A single 4-state 
  variable with memory can create a bias in the tendency to continue
  transitioning in the same direction as in the previous transition, versus going back to the
  previous state. (d) In a multiple spin system, the dynamics can depend on the history of other spins, illustrated by 10 Ising spins arranged in a loop. } \label{fig:toy_model_works}
\end{figure}

We now illustrate the range of possible dynamics generated by GGD, by simulating simple toy models of Ising or Potts spins (see Material and Methods for details on the simulations).
We start by asking whether our illustrative example of a single spin variable with memory can generate non-Markovian tails of the waiting time distribution, as observed in the mice experiments (Fig.~\ref{fig_transition}(b)). 
We define a GGD for a single spin with an exponentially decaying memory kernel,
$\Gamma(t) = A \exp(-t/\tau)$,
where $\tau$ is the time scale of the self-memory, and compare to the classical Glauber dynamics ($A=0$). By construction, both dynamics predict the same steady state distribution, characterized by $\langle \sigma\rangle=\tanh(h)$ (see \sifigsingleisingmh).
However, the GGD predicts a long tail in the waiting time distribution, whereas the naive Glauber dynamics yields an  exponential distribution of waiting times (Fig.~\ref{fig:toy_model_works}(b)). Thus, even a simple form of the memory kernel can create long memory effects similar to those observed in data.

Second, we illustrate the model's ability to account for non-Markovian flow between states.
In the Eco-HAB data, different mice have different levels of roaming---the ratio of probabilities of moving forward versus moving backward in two consecutive transitions. This effect is non-Markovian and arise{\RV{s}} from memory effects. In the GGD, this memory can be encoded in non-diagonal elements of the $4\times 4$ matrix,  $\Gamma(t)$.
We define a GGD on a 4-state variable through a kernel $\Gamma(t)$ with cyclic symmetry and exponentially decaying terms (see detailed parametrization in \sisecggdsinglepotts). 
The model can generate a bias {\RV{between}} transitioning {\RV{to}} the next state (A$\to$ B$\to$ C) rather than coming back to the previous one (A$\to$ B$\to$ A), and this tendency is tuned by the strength $c$ of the off-diagonal elements of the memory kernel {\RV{with a maximum value of $P(A \to B \to C ) = 0.5$}} (Fig.~\ref{fig:toy_model_works}(c), {\RV{\sisecggdsinglepotts}}). 

Mice tend to chase each other in the Eco-HAB experiment ({\sifigmicelf}(c)), 
suggesting that transitions also depend on the history of other mice's behavior. This can be encoded in the GDD through cross-individual memory kernels, $G_{ij}(t)$.
This memory coupling enforces the flipping rate of a degree of freedom (in more general terms, called the follower) to depend on the recent transition of another one (called the leader), such that the transition rate of the leader-follower pairs has a distinguished characteristic timescale that is not visible in other pairs (Fig.~\ref{fig:toy_model_works}(d), see {\sisecggdmultiising} for the memory kernel). 
The symmetry of the memory kernel enforces the memory dependence between a given pair to be symmetric.

\subsection{Inference}
How do we fit the GGD to data?
Assuming that a maximum entropy distribution has already been learned,
we need to solve the inverse problem of finding the memory kernel $\Gamma$ that reproduces the experimentally observed dynamics. We assume an exponential form for the kernel, $\Gamma_i(t)=Ae^{-t/\tau}$, which allows for rewriting $\xi_i(t)$ as an Ornstein-Uhlenbeck process. We maximize the likelihood of the discretized data series $\sigma(t)$ (with some small time bin) over the three parameters $\theta=(\mu,A,\tau)$, using the Expectation-Maximization (EM) algorithm{\RV{~\cite{em_1977}}} 
to deal with the hidden variables $\xi_i(t)$ (see SI). {\RV{Specifically, we adopt the EM algorithm used to infer neural firing dynamics with hidden noise~\cite{smith_brown_2003, kulkarni_paninski_2007, yuan_2010, vroylandt2022likelihood}.}} {\RV{The key difference is that for our inference problem of the GGD, the non-Markovian fluctuation-dissipation relation (Eq.~\ref{eq:fdt_single_ising}) acts as an extra constraint between the parameters generating the noise $\xi(t)$ and the memory kernel $\Gamma_i(t)$, while studies applying EM algorithm to infer neural firing dynamics do not assume the fluctuation-dissipation relation. In addition, the transition dynamics are of the Glauber form (i.e. logistic function, instead of exponential as in models of neural spiking dynamics), which does not lead to simple mathematical expressions and requires Monte-Carlo sampling in the computation (see SI). }}

\begin{figure}[t!]
\centering
\includegraphics[width=\columnwidth]{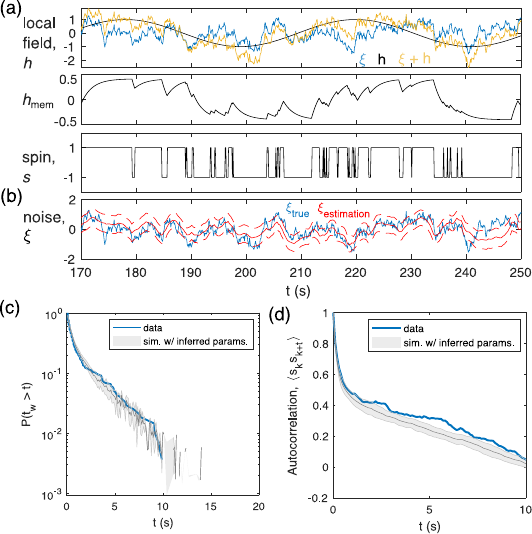}
\caption{Example of GGD inference with an expectation-maximization (EM) algorithm on a single Ising spin under an oscillating field $h(t) = \sin(0.15t)$.  The memory kernel is $\Gamma(t) = A \exp(-t/\tau)$, with true parameters $A = 0.8$, $\tau = 19.5$s, and the baseline transition rate $\mu = 4\text{s}^{-1}$. {\RV{The simulation was conducted for a duration of $T = 300$s, using a time step of $\Delta t = 0.1$s.}}
  (a) the combination of $h(t)$, the noise $\xi$, and the effective
  local field due to history of spin flips, $h_\text{mem}$ (the second
  and third terms of Eq.~\ref{eq:heff_single_ising}), generates the sample spin trajectory $s_t$. (b) Given the spin trajectory $\sigma(t)$ and $h(t)$, the EM algorithm recovers an ensemble of possible realizations of the hidden noise $\xi(t)$.
  With parameters inferred using the EM algorithm, simulated trajectories recovers the waiting time distribution (c) and the autocorrelation decay (d) as the data. The envelop{\RV{e}} of the curve is the standard deviation.
  }\label{fig:ggd_em}
\end{figure}

Figure~\ref{fig:ggd_em}(a) shows an example of a single Ising spin undergoing GGD with a single exponentially-decaying memory kernel. To mimic the situation of a spin within a large interacting system subject to a changing external field $h_i(t)$, we consider a time-dependent field $h(t)$ with sinusoidal form. With our EM algorithm, we are able to recover the parameters with high accuracy
(see {\sifigemparams} and {\sitable}),
as well as estimate the hidden noise (see Fig.~\ref{fig:ggd_em}(b)). 
Trajectories simulated with the inferred set of parameters reproduce the properties of the waiting time distribution (Fig.~\ref{fig:ggd_em}(c)) and the autocorrelation function (Fig.~\ref{fig:ggd_em}(d)) observed in the data. 
{\RV{The error of the EM algorithm, measured as the percentage difference between the EM inferred parameters and the ground truth parameters, scales with data length $T$ as $\sim T^{-1/2}$. This scaling is expected from the Cramer-Rao bound (\sifigemvst, \sifigemvsks(a), top right), and therefore the same as one expects from maximum caliber methods and generalized linear models. }}

We can speed up the inference by heuristically minimizing the distance between the empirical and simulated distributions of dynamical variables, which is defined as the sum of the area between the empirical and model-simulated waiting time cumulative distributions in double logarithmic scale.
Although the error in the inferred parameters is larger than when using EM (\sitable),
the waiting time distribution and the autocorrelation function are recovered correctly (\sifigemks). 

{\RV{One can extend the  parameterization of memory kernel to sums of exponential decays, $\Gamma(t) = \sum_l A_l e^{-t/\tau_l}$, to approximate more general forms of memory kernels which decays at infinite time  (Prony's series, also see~\cite{vroylandt2022likelihood}). The extension of the EM algorithm is straightforward, as the noise can be written as a linear sum of Ornstein-Uhlenbeck processes. }}
{\RV{For the Eco-HAB mice data, we only used the heuristic method, because the EM algorithm becomes unreliable and hard to converge for categorical data with more possible states (see \sifigemvsks ~and \sisecemheu), due to a distortion of the optimization landscape which leads to problems in convergence, which is consistent with the literature (see Chapter 3.4 in~\cite{yuan_thesis}).}}

\subsection{GGD of social mice}
We now go back to our original problem of 15 mice living in an Eco-HAB and ask if we can distinguish properties of individual animals from emergent behavior resulting form interactions. 
Atop the static maximum entropy model we learned previously, we learn the GGD model to fit the waiting time distribution. Since the three non-Markovian features (self memory, individual-specific inertia, and chaser-chased dynamics) occur on different time scales, in principle we should construct a memory kernel whose diagonal and off-diagonal terms have very different time scales. To simplify the task, in addition to the activity prefactors $\mu_i$, we only learn the self-memory kernels $\Gamma_i(t)=A_ie^{-t/\tau_i}$, as this memory occurs on the longest time scale and contributes the most to the observed fat tail in the waiting time distribution. 
Recall from Fig.~\ref{fig_micedata}F that for all mice the waiting time distribution collapses after we divide by its mean, so we can further reduce the number of parameters by assuming $A_i=A_0$, and $\mu_i\tau_i=\textrm{const}$.
We learn this reduced set of dynamical parameters by minimizing the total distance between the observed and predicted waiting time distribution for all mice, computed independently for each mouse while fixing the trajectories of other mice to their experimental values. The optimized parameters {\RV{are}} found to be $A_i = 2.75, \mu_i = 0.25\pm 0.08\text{s}^{-1}, \tau_i = 22 \pm 9\text{s}$.

\begin{figure}
\centering
\includegraphics[width=\columnwidth]{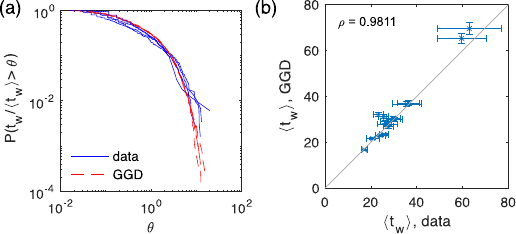}
\caption{The Generalized Glauber dynamics on top of the static maximum entropy model on Eco-HAB data reproduces the long tail of both the shape (panel (a)) and the mean (panel (b)) of the waiting time distributions. {\RV{The Pearson correlation coefficient $\rho$ is given in the plot.}} }\label{fig_mice_GGD}
\end{figure}

We then simulate the dynamics for all 15 mice, using the static parameters learned by pairwise maximum entropy model ($h_{i,\alpha}, J_{ij}$), and the dynamical parameters learned by GGD ($\mu_i, A_i, \tau_i$). 
Simulations correctly capture the tails of the waiting time distribution (Fig.~\ref{fig_mice_GGD}(a) and (b)). By construction, the GGD model reproduces the static observables (\sifigggdmice(a)){\RV{, which the GLM model fails to reproduce (\sifigggdmice(b))}}. 
Since the memory kernel consists of single exponential decays, it suggests a biologically plausible mechanism for its encoding by mice using an iterative leaky integrator of their internal state, without the need to remember their entire past behavior.

{\RV{As discussed in previous paragraphs, while the GGD for Potts spins can tune the probability of consecutive forward transitions within a certain range, the ``roaming'' effect exhibited by the Eco-HAB mice is more pronounced than the GGD allows for in its current form. Specifically, while GGD for a single Potts spin allows a maximum value of $P(A\to B \to C) = 0.5$, in the Eco-HAB mice data, $P(A\to B \to C)_\text{mice}$ is almost always greater than 0.5. }}
 
\section{Discussion}
Motivated by data from the Eco-HAB experiment monitoring the spontaneous collective behavior of mice, we developed an inference method that simplifies dynamical inference by separating the learning of the steady state distribution and the learning of the dynamics. Steady-state inference is performed with the well-developed method of maximum entropy models. Then, a family of reversible dynamics is constructed by adding both a memory kernel and a colored noise, which are related through a non-Markovian fluctuation-dissipation theorem.

GGD has several advantages compared to existing methods of dynamical inference. Unlike maximum caliber methods, it is defined by an explicit form of the transition matrix, making it easier to simulate and interpret. 
{\RV{In addition, maximum caliber methods demand learning of many more parameters. For a system of size $N$, for each time delay of $\Delta t$, maximum caliber needs to simultaneously fit to $N(N-1)/2$ equal time correlations, and the $N(N-1)$ cross-time correlations. A total time lag of $L$ will lead to a total of $N(N-1)/2 + LN(N-1)$ parameters to be fitted simultaneously.  In comparison, in GGD, we first fit $N(N-1)/2$ equal time correlations in the maximum entropy learning step, then for each of the $N$ components, we fit the dynamics to a parameterized memory kernel with a chosen number of parameters (3 in the case we considered).  By separating the learning procedure into first learning the static distribution, and then independently the dynamics for each component, we have much fewer parameters to learn.}}
Unlike generalized linear models, it is guaranteed to agree with the empirical steady-state distribution, and is immune to problems of blow-up divergences that plague inferred state-transition models (see~\cite{park2013spectral, hocker2017multistep, gerhard2017stability, mahuas2020new}, {\RV{also SI Fig. S11}}). Thus, GGD introduces memory while guaranteeing the steady-state. 

{\RV{GGD is a special case of the Generalized Master Equation (GME), as the transition rate is a function of the current state and past states through a memory kernel, and latent variables can be encoded in autoregressively-generated noise in the transition rate. Compared to other GME models, the key ingredient of the GGD is the non-Markovian fluctuation-dissipation relation between the correlation function of the colored noise and the memory kernel, which is essential in restoring the same steady state distribution for a wide range of possible dynamics. }}
The choice of the memory kernels can be very general and is only constrained by the reversibility constraint imposed by the fluctuation-dissipation relation. 
The GGD can capture rich dynamics, such as the effect of individual memory, inertia on discrete dynamics, and 
dynamics which depends on other individuals in the system. While we have focused on continuous-time Glauber dynamics for concreteness, the approach can be extended to any continuous or discrete dynamics (e.g. Metropolis-Hastings). Possible extensions and future applications of GGD include inferring dynamical models for interacting spiking neurons.

We take note of GGD limitations. Its dynamics are reversible by construction, precluding out-of-equilibirum effects. {\RV{On the Eco-HAB data, a GLM model that does not enforce detailed balance shows asymmetricity in the inferred interaction matrix (\sifigglmsmeparam(d)), implying that the Eco-HAB mice are indeed out of equilibrium. Nonetheless, we find a strong correlation between the GLM interaction matrix and the GGD interaction matrix, with correlation coefficient of 0.70 for the asymmetric GLM interaction matrix, and 0.68 for the symmetrized GLM interaction matrix (\sifigglmsmeparam(b, c)), and the local fields $h$ also correlate with a coefficient of 0.69 (\sifigglmsmeparam(a)).}}
Because the memory kernel $\Gamma(t)$ is also an autocorrelation function, it cannot take arbitrary forms. For example, an abrupt suppression to simulate refractory period in neurons would not be possible. 
{\RV{In its current form, the constraint on the memory kernel imposes a limit on the maximum ``roaming'' effect possible, which is less than what the Eco-HAB mice exhibit. }}
On the technical side, we did not manage to reliably infer the effect of memory between individuals $G_{ij}(t)$ from the data. This may be due to the large number of parameters to consider, which scales with the number of pairs. Another difficulty is that the memory of histories of other individuals may happen on a shorter time scale than self-memory (as suggested by the chaser-chased dynamics \cite{winiarski2021social}), confusing the inference procedure. Interestingly, the autocorrelation of the box occupation number decays more slowly than predicted by the model, suggesting that these effect may play an important role (\sifigggdmiceautocorr). 
In addition, because the GGD integrates all memory from the past, it is unable to describe a full memory reset.  Finally, unlike the maximum entropy model, GGD is not a minimal construction that one can use to build dynamics with increasing complexity. One possible solution is to consider GGD with memory kernels built using a complete basis of functions. Such an extension is likely to be very useful, since it could capture phenomena on different timescales, which we saw is relevant for behavior in the Eco-HAB.

\section{Materials and Methods}
\subsection*{The Eco-HAB mice data}
The Eco-HAB system and the appropriate data analysis tools are described in~\cite{ecohab}. The particular experiment and data used for analysis and methods development in this manuscript is published in~\cite{winiarski2021social}. 

\subsection*{Compute normalized chasing rate and rate of sequential flip}
In a lead-and-follow pair, e.g. both the chaser-chased pairs in the Eco-HAB system and the sequential spin flips in multiple interacting Ising spins, we compute the normalized following rate. For a given pair of mice (spins) $(i, j)$ and a fixed time period, we count the number of consecutive transitions where mouse $j$ leads mouse $i$ to leave the same box and to enter the same box, separated by a time difference of $\Delta t$. We then divide this count by an expected null value, computed by cyclically shuffling the time series of all mice, to obtain the normalized lead-and-follow rate.

\subsection*{Learning the static maximum entropy model} 
The static maximum entropy model is learned by gradient descent, where at each optimization step, the parameters $J_{ij}$ and $h_{i,\alpha}$ are updated by the difference between the empirical observable and Monte Carlo sampled observables at the current estimation for each parameter~\cite{gt_spin}. The initial condition is for $h_{i,\alpha}$ to be that of the independent model and all $J_{ij}$ set to zero. The stop condition is when the square difference between the data and the simulation is less than the data variation, computed by bootstrapping the data. Monte Carlo sampling of the model and computing the mean and correlation are performed using the UGM MATLAB package~\cite{ugm}, while all other steps are performed by customized MATLAB codes.

\subsection*{Simulate generalized Glauber dynamics}
To simulate generalized Glauber dynamics with a given memory kernel $\Gamma(t)$, we first generate noise $\xi(t)$ whose correlation is $\langle\xi(t)\xi(t')\rangle = \Gamma(t-t')$ using methods of Fourier transforms~\cite{tuckmantel2008digital}. For systems with higher dimensions, the noise is first independently generated in the eigenbasis of the memory kernel, then transformed back to the standard basis. Then, the dynamics is simulated by discretizing time and using parallel updating. The time step is chosen to be small to make sure at most one spin transitions at any given time step.

\subsection*{Inference of dynamical parameters} 
The inference of the dynamical parameters is performed with two methods. The first is an Expectation-Maximization algorithm (EM) developed in details in SI, and implemented with customized MATLAB code. {\RV{In the single Ising spin example, we chose the stopping criterion such that the absolute change in all three parameters ($\mu, a, \sigma_\varepsilon^2$) must be less than a threshold value of 0.01 over the last 100 iterations of the EM algorithm.}} 

Alternatively, heuristic optimizations use {\RV{two consecutive grid searches followed by a Nelder-Mead algorithm provided by the built-in MATLAB function}} {\tt patternsearch} to find the optimum. 

{\RV{

\subsection*{Learning the generalized linear model} 

For comparison with the generalized Glauber dynamics (GGD), we train a genearlized linear model on the Eco-HAB data. The model follows~\cite{pillow2008spatio} and writes the transition probability in $\Delta t$ as
\beqns
P_{i, \alpha \rightarrow \beta} & = &
\frac{\widetilde{W}_{i, \alpha \rightarrow \beta} \Delta t}{Z_{i, \alpha}}, \\
P_{i, \alpha \rightarrow \alpha} & = &
\frac{1}{Z_{i, \alpha}},
\eeqns
where the transition rate is 
\beqs
\begin{split}
\widetilde{W}_{i, \alpha \rightarrow \beta} 
= \widetilde{\mu}_i \exp 
\Big [
\widetilde{h}_{i\beta}  & - \widetilde{h}_{i\alpha} 
- \sum_j \widetilde{J}_{ij} \delta_{\sigma_i  \sigma_j} \\
& + \int_0^t dt' \widetilde{\Gamma}(t') \sigma_i(t') 
\Big ]
\end{split}
\eeqs
for $\alpha \neq \beta$ with a normalization factor
\beqs
Z_{i,\alpha} = 1 +  \sum_{\beta \neq \alpha, \beta \text{ accessible}} \widetilde{W}_{i, \alpha \rightarrow \beta} \Delta t.
\eeqs
For direct comparison with the GGD model, the memory kernel is chosen to be parameterized by single exponential, $\widetilde{\Gamma}_i(t) = \widetilde{A}_i e^{-t/\widetilde{\tau}_i}$. 
The parameters, $\widetilde{\mu}_i, \widetilde{h}_{i\alpha}, \widetilde{J}_{ij}, \widetilde{A}_i, \widetilde{\tau}_i$ are estimated using maximum likelihood.

 }}

\section*{Acknowledgements}
This work was partially supported by the
European Research Council Consolidator Grant n. 724208, `BRAINCITY - Centre of Excellence for Neural Plasticity and Brain Disorders' project of the Polish Foundation for Science, and the National Science Center grant 2020/39/D/NZ4/01785. The authors are grateful for the discussions and suggestions from Giulio Biroli.

\nocite{*}
\bibliographystyle{pnas}

\onecolumngrid

\newpage

\setcounter{section}{0}
\setcounter{equation}{0}
\renewcommand{\theequation}{S.\arabic{equation}}
\setcounter{figure}{0}
\renewcommand{\theequation}{S\arabic{equation}}
\renewcommand{\thefigure}{S\arabic{figure}}
\renewcommand{\thetable}{S\arabic{table}}

\section*{Supplementary information}
\section{Derivation of generalized Glauber dynamics}

In the main text, we present the generalized Glauber dynamics (GGD) as a method to tune the dynamics of a discrete system, while keeping the steady state distribution fixed. Briefly, we build upon the maximum entropy distribution and ask how we should add an adjustable component to the energy function such that the dynamics can be tuned to fit the data. In this section, we first introduce the standard Glauber dynamics. Then, we present detailed derivations for GGD in three systems with increasing complexities: single binary (Ising) spin with binary states $\pm 1$; multiple binary spins coupled to each other; and single multi-state (Potts) spin with $q$ states.

\subsection{Glauber dynamics satisfies detailed balance}\label{si:sec:detailed_balance}

For a discrete system with a steady state distribution, 
\beqs
P(\sigma) \propto e^{-H(\sigma)},
\eeqs
where $H(\sigma)$ is the Hamiltonian of the system (in the main text we used $E(\sigma)$), 
a natural way to write down a dynamics that can generate such a steady state distribution, is the equilibrium dynamics that satisfies detailed balance. Namely, the transition rate between state $\sigma_\alpha$ and state $\sigma_\beta$ must satisfy
\beqs
W(\sigma_\alpha \rightarrow \sigma_\beta) P(\sigma_\alpha) = W(\sigma_\beta \rightarrow \sigma_\alpha) P(\sigma_\beta). 
\eeqs
The Glauber dynamics is one type of dynamics that obeys detailed balance, where the transition rate is given by
\beqs
W(\sigma_\alpha \rightarrow \sigma_\beta) = \mu \frac{e^{-(H(\sigma_\beta) - H(\sigma_\alpha))}}{1+e^{-(H(\sigma_\beta) - H(\sigma_\alpha))}}
= \mu 
\frac{e^{H(\sigma_\alpha)}}{e^{H(\sigma_\alpha)}+e^{H(\sigma_\beta)}}.
\eeqs

If the Hamiltonian can be expressed by an effective local field, 
\beqs
H(\sigma) = - h \sigma,
\eeqs
the transition rate is
\beqs
W(\sigma_\alpha \rightarrow \sigma_\beta) 
= \mu \frac{e^{-h\sigma_\alpha}}{e^{-h \sigma_\alpha} + e^{-h \sigma_\beta}}.
\eeqs
The effective local field $h$ at time $t$ determines the transition probability in the time window $[t, t+\Delta t)$.

While Glauber dynamics satisfies detailed balance and has a steady state equal to the one given by the Boltzmann distribution, it is Markovian, i.e. the transition rate only depends on the current state of the spins. To include memory in the dynamics, we take inspiration from the multi-dimensional Markov system with equilibrium dynamics, such as hidden Markov models, and generalized Langevin equations. In this case, if we only observe part of the system, the system behaves as if there is memory. Specifically, following the derivation of generalized Langevin equations, we consider coupling the spins to harmonic oscillator baths~\cite{zwanzig_book}. 

For concreteness, we choose to focus on Glauber dynamics, although the method applies to all types of equilibrium dynamics. 

\subsection{GGD for a single Ising spin}\label{si:sec:ggd_single_ising}

Consider a system with a single Ising spin, $\sigma \in \lbrace \pm 1 \rbrace$, described by the Hamiltonian 
\beqs
H_s = - h \sigma.
\eeqs
 The Glauber dynamics transition rate is
 \beq\label{eq:si:w_ggd_single_ising}
W(\sigma \rightarrow -\sigma) = \mu \frac{e^{-h \sigma}}{e^{-h} + e^{h}}.
 \eeq

To include memory, we couple the spin quadratically to a bath of harmonic oscillators indexed by $k$ with momentum $p_k$ and position $q_k$. 
The energy of the system is a sum of the spin energy $H_s$ and the oscillator bath part energy $H_b$, with
\beqns
H_s & = &  -  h \sigma , \\
H_b  & = &  \sum_k \frac{p_k^2}{2} + \frac{1}{2} \omega_k^2 
\left(
q_k - \frac{\gamma_k}{\omega_k^2} \sigma
\right)^2,
\eeqns
with the coefficient $h_s$ being the steady state local field of the spin. Among the coefficients for the oscillators, $\omega_k$ is the frequency of the $k$-th oscillator, and $\gamma_k$ measures the strength of coupling of the system to the $k$-th oscillator. 
We solve for the spin dynamics by integrating out the dynamics of the harmonic oscillators.

Hamilton's equations of motion for the oscillators are 
\beq\label{eq:si:ho_ham_eq_single_ising}
\frac{d q_k}{dt}  =    p_k, 
\hspace{1cm}
\frac{d p_k}{dt}  =   -\omega_k^2  p_k + \gamma_k \sigma.
\eeq
We solve the oscillator dynamics in terms of the spin trajectories,
\beq\label{eq:si:ho_solution_single_ising}
q_k(t) = q_k(0) \cos \omega_k t 
+ p_k(0) \frac{\sin \omega_k t}{\omega_k}
+ \gamma_k \int_0^t
dt' \sigma(t') \frac{\sin \omega_k(t-t')}{\omega_k}.
\eeq
Integration by parts give us
\beq\label{eq:si:ho_solution_single_ising_ibp}
q_k(t) - \frac{\gamma_k}{\omega_k^2}\sigma(t) = A + B
\eeq
where we recognize a term that only depends on the initial condition of the system,
\beqs
A = \left( q_k(0) - \frac{\gamma_k}{\omega_k^2}\sigma(0) \right)
 \cos \omega_k t + p_k(0) \frac{\sin \omega_k t}{\omega_k},
\eeqs
and a term that depends only on the history of the evolution of the spins,
\beqs
B = - \gamma_k \int_0^t
dt' \frac{d\sigma(t')}{dt} \frac{\cos \omega_k(t-t')}{\omega_k^2}.
\eeqs 

Meanwhile, assuming we know dynamics of the harmonic oscillators, we can also write down an effective energy for the Ising spins, integrating out possible configurations of the bath. 

Because the spin obeys $\sigma^2 = 1$, the spin-related part of the energy can be written as
\beqs
H_\text{part.} = - h \sigma - \sum_k \gamma_k  q_k \sigma.
\eeqs
In the limit of infinitesimal time discretization,  we can write down an effective local field for the spin as
\beqs
h_\text{eff} =  h  + \sum_k \gamma_k q_k.
\eeqs

Now, we plug the solution of the harmonic oscillators (\eqref{eq:si:ho_solution_single_ising_ibp}) into the positions $\bm q_k$ for the above effective local fields. With some linear algebra, we can write 
\beq \label{eq:si:heff_oscillators_single_ising}
h_\text{eff} = h + h^\text{bulk}_\text{mem} + h^\text{bd}_\text{mem}  + h_\text{eff}^0(\lbrace x_k(0), v_k(0)\rbrace).
\eeq

We can analyze this effective local field term by term. The term involving the history of the spins is
\beq\label{eq:si:heff_mem_single_ising}
\begin{split}
h^\text{bulk}_\text{mem} 
= & - \int_0^t dt' 
\sum_k \frac{\gamma_k^2}{\omega_k^2}
\cos \omega_k (t-t') \frac{d \sigma_\alpha(t')}{dt'} \\
= & -  \int_0^t dt' \Gamma (t-t') \frac{d  \sigma(t')}{dt'}.
\end{split}
\eeq
From the first to the second line, we recognize the sum over the harmonic oscillators as a cosine Fourier series.

The boundary term comes from integration by parts, and is given by the left hand side of \eqref{eq:si:ho_solution_single_ising_ibp}, 
\beqs
h^\text{bd}_\text{mem} 
= \sum_k \frac{\gamma_k^2}{\omega_k^2}  \sigma(t) = \Gamma(0) \sigma(t).
\eeqs

The term that involves the initial position of the oscillators, $h_\text{eff}^0 (\lbrace q_k(0), p_k(0)\rbrace)$, constitutes the noise in the dynamics.  We replace the initial conditions with their average values, which are stochastic variables with colored noise,
\beq
\xi(t) \equiv h_\text{eff}^0 (\lbrace x_k(0),  v_k(0)\rbrace) = \sum_k \gamma_k \left[
\left( q_k(0) - \frac{\gamma_k}{\omega_k^2} \sigma(0) \right) \cos \omega_k t
+ p_k(0) \frac{\sin \omega_k t}{\omega_k}
\right].
\eeq
Based on the equipartition theorem, the expectation values for initial position at temperature $k_B T = 1$ is
\beqs
\left\langle \left[q_k(0) - \frac{\gamma_k}{\omega_k^2} \sigma(0) \right]^2 \right\rangle = 
\frac{1}{\omega_k^2},
\hspace{1cm}
\left\langle p_k(0)^2 \right\rangle = 1.
\eeqs
After some algebra, we find the noise correlation as
\beq\label{eq:si:noise_corr_single_ising}
\begin{split}
\langle \xi(t) \xi(t') \rangle 
& = \sum_k \frac{\gamma_k^2}{\omega_k^2}
(\cos \omega_k t \cos \omega_k t' + 
\sin \omega_k t \sin \omega_k t') \\
&= \sum_k \dfrac{\gamma_k^2}{\omega_k^2}\cos \omega_k (t-t')\\
   &= \Gamma(t - t').
 \end{split}
\eeq

In summary, the effective local field is 
\beq\label{eq:si:heff_single_ising}
h_\text{eff} =  h
+ \Gamma(0) \sigma(t) 
- \int_0^t dt' \Gamma (t-t') \frac{d \sigma(t')}{dt'} + \xi(t),
\eeq
where the noise correlation satisfies \eqref{eq:si:noise_corr_single_ising}. The first term is the static local field, as learned from the steady state model, the second and the third terms describe the memory, and the last term the colored noise that results from the coupling from the memory kernel. The memory term depends on the coupling of the initial value of the memory function and later values of the spins (the second term) as well as the convolution of the memory kernel and the spin flipping rate. For clarity in the main text, we denote the second and the third term as
\beqs
h_\text{mem} = \Gamma(0) \sigma(t) - \int_0^t dt' \Gamma (t-t') \frac{d \sigma(t')}{dt'}.
\eeqs

The transition rate of the spin is given by
\beq\label{eq:si:w_ggd_single_ising}
W(\sigma \rightarrow -\sigma) = \mu \frac{e^{-h_\text{eff} \sigma}}{e^{-h_\text{eff}} + e^{h_\text{eff}}}. 
\eeq
Notice that the mathematical form of the transition rate is identical to that of the naive Glauber dynamics (\eqref{eq:si:w_ggd_single_ising}).  
Nonetheless, the effective local field $h_\text{eff}$ depends on the history of spin $\sigma$, which adds a memory dependence to the dynamics. By construction, detailed balance is ensured in the whole system that includes both the spin and oscillator baths. 

\subsection{GGD for multiple Ising spins}\label{si:sec:ggd_multi_ising}

Now we increase the complexity of the generalized Glauber dynamics, and consider a system of $N$ interacting binary (Ising) spins, $\bm\sigma^\intercal = (\sigma_1, \sigma_2, \dots, \sigma_N)$.  
For a  small enough time interval, at most one spin flips at a time, so the Glauber dynamics is given by the same form as in the single Ising system, 
\beq
W(\bm \sigma = \left(
\sigma_1, \dots, \sigma_i, \dots, \sigma_N)^\intercal
\rightarrow 
\bm \sigma' =
(\sigma_1, \dots, -\sigma_i, \dots, \sigma_N)^\intercal
\right)
= 
\mu_i \frac{e^{-h_i \sigma_i}}{e^{-h_i} + e^{h_i}} 
= 
\mu_i \frac{e^{-\bm h^\intercal \bm \sigma}}{e^{-\bm h^\intercal \bm \sigma} + 
e^{-\bm h^\intercal \bm \sigma'}}.
\eeq

Now we couple each spin $\sigma_i$ quadratically to a bath of harmonic oscillators with momentum $p^{(i)}_k$ and position $q^{(i)}_k$. 
The energy of the system is a sum of the spin energy $H_s$ and the oscillator bath part energy $H_s$, with
\beqn
H_s & = &  - \bm g^\intercal \bm\sigma  - \frac{1}{2} \bm\sigma^\intercal \bm J \bm\sigma \equiv -\bm h^\intercal \bm\sigma , \\
H_b  & = &  \sum_k \frac{\bm{p}_k^\intercal \bm{p}_k}{2} + \frac{1}{2} \omega_k^2 
\left(
\bm{q}_k - \frac{\gamma_k}{\omega_k^2}  \bm\sigma
\right)^\intercal 
\bm M
\left(
\bm{q}_k - \frac{\gamma_k}{\omega_k^2} \bm\sigma
\right), \label{eq:si:ham_bath_multiple_ising}
\eeqn
where the coefficients $\bm{g}$ are the local fields for the spins, $\bm J$ the pairwise interactions among the spins, and 
$\bm h  = \bm g + \bm J \bm \sigma$ the steady state local field. Among the coefficients for the oscillators, $\omega_k$ is the frequency for the $k$-th oscillator, and $\gamma_k$ measures the strength of coupling of the system to the $k$-th oscillator. For convenience, we set identical coupling strength for each spin $\sigma_i$. For simplicity, we assume that the coupling matrix among the baths, $\bm M$, is symmetric and positive definite, $\bm M = \bm O \bm D \bm O^{-1}$, where  $O$ is a matrix of eigenbases of $M$. We denote $\bm O^{-1} \bm q_k = \bm x_k$, $\bm O^{-1} \bm p_k = \bm v_k$, $\bm O^{-1} \bm \sigma  = \widetilde{\bm \sigma} $, and the eigenvalues of the matrix $\bm M$ using $\lambda_\alpha$. 

In the eigenbasis of $M$, Hamilton's equations of motion for the oscillators are 
\beqs
\frac{d\bm x_k}{dt}  =   \bm v_k, 
\hspace{1cm}
\frac{d \bm v_k}{dt}  =   -\omega_k^2 \bm D \bm v_k + \gamma_k \bm D \widetilde{\bm \sigma}.
\eeqs 
This set of equation is analogous to \eqref{eq:si:ho_ham_eq_single_ising} and can be solved accordingly. 

We solve the oscillator dynamics in terms of the spin trajectories,
\beq\label{eq:si:ho_solution}
x_{k,\alpha}(t) - \frac{\gamma_k}{\omega_k^2}\widetilde{\sigma}_\alpha(t) =  A + B,
\eeq 
where we recognize a term that only depends on the initial condition of the system,
\beqs
A = 
\left(
x_{k,\alpha}(0) -  \frac{\gamma_k}{\omega_k^2} \widetilde{\sigma}_\alpha(0)
\right)
\cos \omega_k \lambda_\alpha^{1/2} t + v_{k,\alpha}(0) \frac{\sin \omega_k \lambda_\alpha^{1/2} t}{\omega_k \lambda_\alpha^{1/2}},
\eeqs
and a term that depends only on the history of evolution of the spins,
\beqs
B = 
- \gamma_k \lambda_\alpha \int_0^t dt' \frac{d\widetilde{\sigma}_\alpha(t')}{dt} \frac{\cos \omega_k \lambda_\alpha^{1/2} (t-t')}{\omega_k^2 \lambda_\alpha} .
\eeqs 

Meanwhile, for the Ising spins, we can write down an effective energy that depends on the harmonic oscillators.
The spin-related part of the energy is
\beq\label{eq:si:hpart_multi_ising}
H_\text{part.} = -\bm h^\intercal \bm \sigma + \sum_k  \frac{1}{2} \frac{\gamma_k^2 }{\omega_k^2}\bm \sigma^\intercal \bm M \bm \sigma - \gamma_k \bm q_k^\intercal \bm M \bm \sigma.
\eeq
In the limit of the infinitesimal time discretization, at each time point there is maximally a single spin flip, we can write down an effective local field for the spins such that the difference of the Hamiltonian between two states $\bm \sigma, \bm \sigma'$, separated by a single spin flip is equal to $ \bm h_\text{eff}^\intercal \bm \sigma - \bm h_\text{eff}^\intercal\bm \sigma' $. The effective local field is
\beqs
\bm h_\text{eff} = \bm h -  \left[
\sum_k \frac{\gamma_k^2}{\omega_k^2} \bm M 
- \text{diag}\left(
\sum_k \frac{\gamma_k^2}{\omega_k^2} \bm M 
\right)
\right] \bm \sigma  
+ \gamma_k \bm M \bm q_k.
\eeqs
Here, the symbol $\text{diag}(\bm A)$ indicates the diagonal matrix of matrix $\bm A$, with the $i,j-$th entry being $\delta_{ij}A_{ij}$. 

Now, we can plug the solution of the harmonic oscillators into the positions $\bm q_k$ for the above effective local fields. With some linear algebra, we can write 
\beq \label{eq:si:heff_multi_ising_oscillators}
\bm h_\text{eff} = \bm h
- \left[ \sum_k \frac{\gamma_k^2}{\omega_k^2} \bm M -
\text{diag}\left(
\sum_k \frac{\gamma_k^2}{\omega_k^2} \bm M
\right) \right] \bm\sigma 
+ \bm h^\text{bulk}_\text{mem} 
+ \bm h^\text{bd}_\text{mem}  
+ \bm h_\text{eff}^0(\lbrace x_k(0), v_k(0)\rbrace).
\eeq

We can analyze this effective local field term by term. The term involving the history of the spins is
\beq\label{eq:si:heff_mem}
\begin{split}
\bm h^\text{bulk}_\text{mem} 
= & - \bm O  
\int_0^t dt' 
\left(
\lambda_\alpha \sum_k \frac{\gamma_k^2}{\omega_k^2}
\cos \omega_k \lambda_\alpha^{1/2} (t-t') \frac{d\widetilde{\sigma}_\alpha(t')}{dt'}
\right) \\
= & - \bm O \int_0^t dt' \widetilde{\bm \Gamma}(t-t') \frac{d\widetilde{\bm \sigma}(t')}{dt'} \\
= & -  \int_0^t dt' \bm \Gamma (t-t') \frac{d\bm \sigma(t')}{dt'}.
\end{split}
\eeq
From the first to the second line, we recognize the sum over the harmonic oscillators as a cosine Fourier series, and define $\widetilde{\bm \Gamma}(t)$ to be a diagonal matrix, with the diagonal entries as
\beqs\label{eq:si:mem_kernel_single_potts_tilde}
\widetilde{\bm \Gamma}_{\alpha\alpha}(t) \equiv 
\lambda_\alpha
\sum_j \dfrac{\gamma_j^2}{\omega_j^2}\cos \omega_j \lambda_\alpha^{1/2}t
\eeqs
and 
\beqs\label{eq:si:mem_kernel_single_potts}
\bm \Gamma(t) \equiv \bm O \widetilde{\bm \Gamma}(t) \bm O^{-1}.
\eeqs

The boundary term comes from integration by parts,  
\beqs
\bm h^\text{bd}_\text{mem} 
= \sum_k \frac{\gamma_k^2}{\omega_k^2} \bm M \bm \sigma(t) = \bm \Gamma(0) \bm \sigma(t).
\eeqs
Notice that similar terms appear in the first half of the effective local field (\eqref{eq:si:heff_multi_ising_oscillators}).

The term that involves the initial position of the oscillators, $\bm h_\text{eff}^0 (\lbrace \bm x_k(0), \bm v_k(0)\rbrace)$, constitutes the noise in the dynamics. Based on the equipartition theorem, we replace the initial conditions with their average values, which are stochastic variables with colored noise,
\beq
\bm \xi(t) = \bm h_\text{eff}^0 (\lbrace \bm x_k(0), \bm v_k(0)\rbrace).
\eeq
After some algebra, we find the noise correlation as
\beq\label{eq:si:noise_corr_multi_ising}
\begin{split}
\langle \bm \xi(t) \bm \xi(t')^\intercal \rangle &= 
\bm O 
\left(
\begin{matrix} \lambda_\alpha
\sum_j \dfrac{\gamma_j^2}{\omega_j^2}\cos \omega_j \lambda_\alpha^{1/2}(t-t')
\end{matrix}
\right)
 \bm O^{-1} \\
   &= \bm \Gamma(t - t')
 \end{split}
\eeq

In summary, the effective local field is 
\beq\label{eq:si:heff_multi_ising}
\begin{split}
\bm h_\text{eff} & = \bm h
- \left[ \bm\Gamma(0)
- \text{diag} \left( \bm \Gamma(0) \right)
\right] \bm\sigma(t) 
+ \bm\Gamma(0)\bm\sigma(t) - \int_0^t dt' \bm \Gamma (t-t') \frac{d\bm\sigma(t')}{dt'} + \bm\xi(t),\\
& = \bm h
+ \text{diag} \left( \bm \Gamma(0) \right)
\bm\sigma(t) 
- \int_0^t dt' \bm \Gamma (t-t') \frac{d\bm\sigma(t')}{dt'} + \bm\xi(t),
\end{split}
\eeq
where the noise correlation satisfies \eqref{eq:si:noise_corr_multi_ising}. The first term is the static local field, as learned from the steady state model, the second and the third terms describe the memory, and the last term the colored noise that results from the coupling from the memory kernel. The memory term depends on the coupling of the initial value of the memory function and later values of the spins (the second term) as well as the convolution of the memory kernel and the spin flipping rate. We can recognize the diagonal elements of the memory kernel $\Gamma_i(t) \equiv \Gamma_{ii}(t)$ as terms for self-memory, i.e. how the dynamics of the spin is coupled to its own history. The off-diagonal terms, $G_{ij}(t) \equiv \Gamma_{ij}(t)$ for $i \neq j$, couples the transition probability of a given spin to the history of other spins.

For Fig.~3(d) from the main text, we generate the dynamics of 10 Ising spins in a loop, by setting the local field $\bm h_s = 0$ such that the equal-time correlation between different spins are zero, and the memory kernel generated by
\beqs
\Gamma_{ij}(0) = 
\begin{cases}
1, & i = j, \\
-0.4, & \lvert i - j \rvert = 1, \\
0, & \text{otherwise}.
\end{cases}
\eeqs 
 We diagonalize $\bm \Gamma(0)$ to obtain the orthogonal basis $\bm O$ and the eigenvalues $\lambda_\alpha$, and we
set the memory kernel at larger time as
\beqs
\bm \Gamma(t-t') = \bm O \left( 
\lambda_\alpha e^{-\lambda_\alpha^{1/2} t/\tau} 
\right) \bm O^{-1}.
\eeqs

\subsection{GGD for a single Potts spin}\label{si:sec:ggd_single_potts}

We now extend it to single Potts spin with $\rho$ states, where we denote the states using unit vectors in the standard basis. For example, for the Eco-HAB mice dataset, where the mice have $\rho = 4$ distinct states, the spin takes possible values of 
\beqs
\bm \sigma \in \left\lbrace
\left( \begin{matrix}
1\\0\\0\\0
\end{matrix}\right),
\left( \begin{matrix}
0\\1\\0\\0
\end{matrix}\right),
\left( \begin{matrix}
0\\0\\1\\0
\end{matrix}\right),
\left( \begin{matrix}
0\\0\\0\\1
\end{matrix}\right)
\right\rbrace.
\eeqs
The Hamiltonian can again be expressed using a local field,
\beqs
H(\bm \sigma) = - \bm h^\intercal \bm \sigma.
\eeqs

\subsubsection{Uncoupled states}
We first consider the case where these states are uncoupled, i.e. the transition rate from a given state only depend{\RV{s}} on histories of the spin in the same state. Mathematically, the Hamiltonian can be decoupled into a sum of terms that are independent for each state, and the effective local field is identical to that of the binary spins,
\beq
\bm h_\text{eff}  = \bm h + \Gamma(0)\bm \sigma(t) - \int_0^t dt' \frac{d\bm \sigma(t')}{dt'}\Gamma(t-t')  + \bm \xi (t).
\eeq
The auto-correlation of the noise is
\beq
\langle \xi_\alpha(t) \xi_\beta(t') \rangle = \Gamma(t-t')\delta_{\alpha\beta}.
\eeq
One can check that in the limit of number of states $\rho = 2$, the solution for single Ising spin with GGD (with parameters denoted by tildes) is recovered by identifying $h^{+} - h^{-} = 2 \widetilde{h}$ and $\Gamma(t) = 2 \widetilde{\Gamma}(t)$.

\subsubsection{Coupled states}~\label{SI:single_potts_states_coupled}

The memory kernel can also couple different Potts states. In this case, the Hamiltonian for the bath of oscillators coupled to $q$ spins is
\beq
H_b   =  \sum_j \frac{\bm{p}_j^\intercal \bm{p}_j}{2} + \frac{1}{2} \omega_j^2 
\left(
\bm{q}_j - \frac{\gamma_j}{\omega_j^2}  \bm\sigma
\right)^\intercal 
\bm M
\left(
\bm{q}_j - \frac{\gamma_j}{\omega_j^2} \bm\sigma
\right).
\eeq
For simplicity, we assume that the coupling matrix $\bm M$ is symmetric and positive definite. 
We notice that this Hamiltonian is identical to the Hamiltonian for multiple interacting Ising spins (\eqref{eq:si:ham_bath_multiple_ising}), and so is the spin-dependent part of the Hamiltonin (\eqref{eq:si:hpart_multi_ising}),
\beqs
H_\text{part.} = -\bm h^\intercal \bm \sigma + \sum_k  \frac{1}{2} \frac{\gamma_k^2 }{\omega_k^2}\bm \sigma^\intercal \bm M \bm \sigma - \gamma_k \bm q_k^\intercal \bm M \bm \sigma.
\eeqs
 The only difference is the values taken by the spin $\bm \sigma$, so the second term of $H_\text{part.}$ takes a different value. The effective local field is then
\beq
\bm h_\text{eff}   =  
\bm h - \sum_j \frac{1}{2}\frac{\gamma_j^2}{\omega_j^2} \text{diag}(\bm M)  + \sum_j \gamma_j \bm M \bm q_j,
\eeq
and, following analogous procedures as in the multiple Ising case, can be further reduced as 
\beq \label{eq:si:heff_single_potts}
\bm h_\text{eff} = \bm h_\text{s} - \frac{1}{2} \text{diag}\left(\bm \Gamma (0)\right) + \bm \Gamma(0) \bm \sigma(t) - \int_0^t dt' \bm \Gamma(t-t')\frac{d\bm \sigma(t')}{dt'} + \bm \xi(t). 
\eeq
The noise correlation satisfies
\beq
\langle \bm \xi(t) \bm\xi(t')^\intercal \rangle = \bm \Gamma(t-t') 
\eeq
In the case that $\bm M = \bm 1_\rho$, we recover the expression for a single Potts spin with memories where the states are uncoupled. Now, the memories are coupled such that the transition rate from a specific state can be effected by the spin history in another state.

In Fig.~3(c) from the main text, we choose the memory kernel to observe the symmetry of the Eco-HAB, by parametrizing the memory kernel 
\beqs
\bm \Gamma(0) = \left(
\begin{matrix}
1 & b & c & b \\
b & 1 & b  & c \\
c & b & 1 & b \\
b & c & b & 1
\end{matrix}
\right) 
\eeqs
and assume it exponentially decays for $t > 0$.
{\RV{ The noise is generated independently in the eigenspace of the memory kernel $\bm \Gamma(0)$, each with a correlation that decays according to a single exponential with the same correlation time $\tau$, then projected back to the original space. 

Because $\bm \Gamma(t)$ is the correlation function of the noise, $\bm \Gamma(0)$ needs to be positive definite, which imposes that $c < 1$ and $\vert b \vert < (1+c)/2$. 

Averaged over the noise $\xi$, and approximating with only the most recent transition, the effective local field is 
\beqns
\langle h_A\rangle & \approx & h_{s, A} - \frac{1}{2} + b - (b-1) e^{-(t-t_0)/\tau}, \\
\langle h_B\rangle & \approx & h_{s, B} + \frac{1}{2} - (1-b) e^{-(t-t_0)/\tau}, \\
\langle h_C\rangle & \approx &h_{s, C} - \frac{1}{2} + b - (b-c) e^{-(t-t_0)/\tau}.
\eeqns

Assuming the simplest case where the static local fields are the same for each state ($h_A = h_C$), we have $\langle h_C \rangle < \langle h_A \rangle$. This means the more likely transition for the next state is $A$, i.e. $P(A \to B \to A) > P(A \to B \to C)$ with the limit of equality achieved at $c \to 1$.

}}

\section{Formulating the noise using Yule-Walker equations}

To ensure that the memory kernel in GGD is a well-defined cross-correlation function, we choose to first generate the hidden variable $\xi_t$, and then find the functional form of the memory kernel. 

We consider the broad class of models, where the hidden variable $\xi_t$ can be generated by linear dynamics
\beq
\bm \xi_t = \sum_{l = 1}^p \bm A_l \bm \xi_{t-l} + \bm \varepsilon_t,
\label{eq:si:var_noise}
\eeq
with the residue correlation $\bm \Sigma = \langle \bm \varepsilon \bm \varepsilon^\intercal \rangle_t$.
Multiplying both sides by $\bm \xi_{t-k}^\intercal$ and averaging over $t$, we find that the auto-covariance sequence 
{\RV{$\bm \Gamma_k = \langle \bm \xi_t \bm \xi_{t-k}^\intercal \rangle_t$}}
follows the Yule-Walker equation
\beqs
\bm \Gamma_k = \sum_{l=1}^p \bm A_l \bm \Gamma_{k-l} + \delta_{0 k}\bm \Sigma.
\eeqs

The simplest form of noise correlation is when the noise is generated with a maximum of 1 time-step coupling to the past (VAR(1)), which gives an Ornstein-Uhlenbeck process. Here, one can solve for $\Gamma$ given the parameters in the noise, the coupling parameters $\bm A$ and the noise correlator $\bm \Sigma$,
\beqs
\begin{split}
\bm \Gamma_0  &  = \bm A \bm \Gamma_1^\intercal + \bm \Sigma,\\
\bm \Gamma_1 & = \bm A \bm \Gamma_0.
\end{split}
\eeqs

In the example of a single Ising spin, the effective local field can be written as ($k\Delta = t$).
\beqs
\begin{split}
h^\text{eff}_k & =  h_k + \Gamma_0 \sigma_k
- \sum_{q=1}^\infty \Gamma_{q} (\sigma_{k-q+1} - \sigma_{k-q})
+ \xi_k \\ 
& = h + \Gamma_0 \sigma_k - \sum_{q=1}^\infty A^q \Gamma_0 
(\sigma_{k-q+1} - \sigma_{k-1}) + \xi_k.
\end{split}
\eeqs
where $\Gamma_0 = \frac{\Sigma}{1 - \bm A^2}$ and $\xi_k = A\xi_{k-1} + \varepsilon_k$.

\section{Inferring dynamical parameters with an expectation-maximization algorithm}

For clarity of the notations for later calculation, from now we use $s$ instead of $\sigma$ for the spin. 
We specify
$s_{0,k} = {s_0, s_1, \dots, s_k}$: the spin configurations from time $t = 0$ to time $t = k\Delta$.  Denoting functions evaluated at $t = k\Delta$ with a subscript $k$, we can write the discretized dynamics as \beq\label{si:eq:discrete_heff}
h_k^\text{eff} = h_k + \Gamma_0 s_k
- \sum^{\infty}_{q=1} \Gamma_q (s_{k-q+1} - s_{k-q}) 
+ \xi_k.
\eeq
This effective local field $h_k^\text{eff}$ determines the transition probability for time $t \in  [k\Delta, (k+1)\Delta)$.

Assuming $\Gamma(\Delta t) = A e^{-\Delta t/\tau}$, we rewrite the noise as generated by an Ornstein-Uhlenback process, and the discrete dynamics of the effective local field (Eq.~(\ref{si:eq:discrete_heff})) become
\beq \label{eq:state_space_model}
\begin{cases}
h_k^\text{eff} 
  =  h_k + A s_k - \sum_{q=1}^\infty A a^q (s_{k-q+1} - s_{k-q} )
+ \xi_k, \vspace{0.3cm}\\
\xi_k  =  a \xi_{k-1} + \varepsilon_k,
\hspace{1cm}
\varepsilon_k \sim \mathcal{N}(0, \sigma_\varepsilon^2). 
\end{cases}
\eeq
The non-Markovian fluctuation-dissipation theorem (\eqref{eq:si:noise_corr_single_ising}) imposes the mathematical relationships $A = \sigma_\varepsilon^2/(1-a^2)$ and $\tau = - 1 / \ln a$. 
These equations are the basis for the EM algorithm.

The coupled dynamical equation in \eqref{eq:state_space_model} is a state-space model, where a hidden variable, the noise $\xi$ stochastically determines the observed dynamics of the spin $s$. 
Inferring parameters for such state-space models with hidden variables has been studied extensively, for continuous dynamics in the context of Kalman filters~\cite{kalman1960new}, and for discrete dynamics in the context of point-process adaptive filters developed in the field of neuroscience~\cite{eden2004dynamic}, typically using an expectation-maximization (EM) algorithm~\cite{em_1977} to perform  maximum likelihood estimation (MLE).
 Following the literature~\cite{smith_brown_2003, kulkarni_paninski_2007, lawhern_paninski_2010, yuan_2010} we will develop an EM algorithm to learn the parameters of GGD. The differences of the GGD model compared to previous work is that 1) the dynamics is Glauber and 2) the parameters for generating the hidden noise $\xi$, and for the memory kernel ($a, s_\varepsilon^2$), need to obey the relations imposed by the non-Markovian fluctuation-dissipation theorem as given by \eqref{eq:si:noise_corr_single_ising}.

Specifically, our goal is to find the set of parameters $\bm\theta = \lbrace \mu, A, \tau \rbrace$ (or equivalently $\widetilde{\bm\theta} = \lbrace \mu, a, \sigma_\varepsilon^2 \rbrace$) that maximize the probability of the data, $\ p(s_{0,K} \vert \bm \theta ) $.

While a direct evaluation over this probability is difficult to obtain, the EM algorithm is an iterative method that finds (local) maxima of the data likelihood, by iteratively calculating the expectation of the log-likelihood of the data evaluated by averaging of the  posterior distribution of the noise based on the current  latent parameter estimates $\theta^{(l)}$, and updating the latent parameters by maximizing the current estimate of the likelihood, $Q(\bm \theta \vert \bm \theta^{(l)})$:

\beq\label{eq:si:q}
\begin{split}
Q(\bm \theta \vert \bm \theta^{(l)})  
&  = \mathbb{E}_{s_{0,K}, \bm{\theta}^{(l)}}[\log p(s_{0,K}, \xi_{0,K} \vert \theta) ] \\
&  = \mathbb{E}_{s_{0,K}, \bm{\theta}^{(l)}}[\log p(s_{0,K} \vert \xi_{0,K}, \mu)]  +
\mathbb{E}_{s_{0,K}, \bm{\theta}^{(l)}}[ \log p(\xi_{0, K} \vert a, \sigma^2_\varepsilon) ]. \\
\end{split}
\eeq
We have rewritten the likelihood two terms. The first term is given by the transition rate function, which we approximate by assuming $\Delta$ is small 
\beqs
\begin{split}
p(s_{0, K}\vert \xi_{0,K}, \mu) 
& = \exp\left \lbrace
\sum_{\text{transition at } k}  \left[
\log W (k\Delta \vert \xi_k, s_{0,k} , \mu) 
+ \log \Delta
\right]
- \sum_{\text{no trans. at } k}^K W (k\Delta \vert \xi_k, s_{0,k}, \mu) \Delta
\right \rbrace \\
&  \approx \exp\left \lbrace
\sum_{k = 0}^K   \mathbf{1}(k\Delta) \left[
\log W(k\Delta \vert \xi_k, s_{0,k}, \mu) 
+ \log \Delta
\right]
- W(k\Delta \vert \xi_k, s_{0,k}, \mu) \Delta
\right \rbrace,
\end{split}
\eeqs
where $I(k\Delta) \equiv \mathbf{1}(s((k+1)\Delta) = - s(k\Delta) )$ is the indicator function, i.e. 1 when there is a transition in time $t \in (k\Delta, (k+1)\Delta]$ and 0 otherwise. {\RV{$W(k\Delta)$ is the state-dependent transition rate for any transitions between time point $k\Delta$ and time point $(k+1)\Delta$, given by Eq.~\ref{eq:si:w_ggd_single_ising}}.} The second term is given by the Gaussian probability of the noise, as specified in \eqref{eq:state_space_model},
\beqs
p(\xi_{0,K} \vert a, \sigma^2_\varepsilon) = 
\left( \frac{1-a^2}{2\pi \sigma^2_\varepsilon} \right)^{1/2} \left( \frac{1}{2\pi \sigma^2_\varepsilon} \right)^{K/2} \exp \left\lbrace
-\frac{1}{2} \left[
\frac{(1-a^2)}{\sigma_\varepsilon^2} \xi_0^2
+ \sum_{k=1}^K \frac{(\xi_k - a \xi_{k-1})^2}{\sigma_\varepsilon^2}
\right]
\right \rbrace.
\eeqs

The EM algorithm finds the optimum of the marginal likelihood by alternating between two steps.  
At each iteration step $l$, the E(xpectation)-step computes $Q(\bm \theta \vert \bm \theta^{(l)})$ with respect to the posterior distribution of the noise, given the current estimate of parameters, $\lbrace \bm \theta^{(l)} \rbrace$.  
The posterior distribution of the noise is approximated by a Gaussian distributions and estimated with a point process estimator and backward Kalman smoothers, which track the one- and two-point statistics.  
Then, in the M(aximization)-step, the complete log likelihood $Q(\bm \theta \vert \bm \theta^{(l)})$ is maximized, and the argument is used as the updated parameter $\bm \theta^{(l+1)}$. 

\subsection{Computing posterior distribution of the noise in the E-step}

At each iteration, the E-step computes the expectation value of $Q(\bm \theta \vert \bm \theta^{(l)})$, the complete likelihood of the data $s_{0, K}$ and the hidden variable, noise $\xi_{0, K}$, over the posterior distribution of $\xi_{0,K}$, given the current estimation of parameters $\bm \theta^{(l)}$. Following the literature~\cite{smith_brown_2003, kulkarni_paninski_2007}, the posterior distribution of $\xi_{0, K}$ is estimated as Gaussian distributions, whose mean and covariance are denoted as 
\begin{align*}
\xi_{k\vert K} & \equiv  
\mathbb{E}_{s_{0,K}, \bm{\theta}^{(l)}}
\left[
\xi_k
\right] \\
\sigma^2_{k \vert K} & \equiv 
\mathbb{E}_{s_{0,K}, \bm{\theta}^{(l)}} 
\left[
\xi_k^2
\right] - \xi_{k\vert K}^2  \\
\sigma_{k,k+1\vert K} & \equiv 
\mathbb{E}_{s_{0,K}, \bm{\theta}^{(l)}}
\left[
\xi_k \xi_{k+1}
\right]
- \xi_{k\vert K} \xi_{k+1 \vert K}.
\end{align*}
The notation $k \vert K$ denotes the point estimates for the noise $\xi$ at the $k$-th time point given the entire observation, $s_{0, K}$. We denote the point estimates given observation up to the $j$-th time point using the subscript $k \vert j$. The mean and covariances of the noise are computed in three steps: first, we use a forward point process filter to compute $\xi_{k\vert k}$ and $\sigma^2_{k \vert k}$; second, a backward Kalman smoother to compute $\xi_{k \vert K}$ and $\sigma^2_{k \vert K}$; and finally, a state-space covariance algorithm to compute $\sigma_{k, k+1\vert K}$. \\

\textit{i. Forward point process filter}

The posterior distribution of the noise is computed recursively. Given the mean and variance of the noise at time point $k-1$ {\RV{and the entire history of spin and spin updates up to time point $k$, $H_k = \lbrace s_{0,k}, ds_k \rbrace$}}, the posterior distribution of the noise at time point $k$ is given by
{\RV{
\beq \label{eq:em:noise_posterior}
\begin{split}
p(\xi_k \vert H_k )
& = \frac{p(\xi_k \vert s_{0,k} ) 
p(ds_{k} \vert \xi_k, s_{0,k})}{p(ds_{k}\vert s_{0,k})} \\
& = \frac{p(ds_{k} \vert \xi_k, s_{0,k})
\int p (\xi_k \vert \xi_{k-1} )
p(\xi_{k-1} \vert s_{0, k}) d \xi_{k-1}
}
{p(ds_{k}\vert s_{0,k})} \\
& \propto \exp \left\lbrace
-\frac{1}{2}\frac{(\xi_k - \xi_{k\vert k-1})^2}
{\sigma^2_{k\vert k-1} }
\right \rbrace
 \exp\left \lbrace
\log W(k\Delta)
\mathbf{I}(k\Delta)  
- W(k\Delta) \Delta
\right \rbrace.
\end{split}
\eeq
}}

Based on Gaussian continuity assumption, we have
\beqs
\xi_{k\vert k-1} = a \xi_{k-1 \vert k-1},
\hspace{1cm} 
\sigma^2_{k \vert k -1 } = 
\sigma_\varepsilon^2 
+ a^2 \sigma^2_{k-1 \vert k-1}.
\eeqs

The mean of the posterior distribution of $\xi_k$ can be estimated using importance Monte Carlo sampling. For $N_\text{MC}$ random numbers $\xi_\text{i}^\text{MC}$ sampled from the Gaussian distribution $\mathcal{N}(\xi_{k\vert k-1}, \sigma^2_{k \vert k-1})$, the Monte Carlo estimation of the mean is
\beq\label{eq:em:mean_noise_mc}
\xi_{k \vert k}^\text{MC} = 
\frac{\sum_{i=1}^{N_\text{MC}}\xi_i^\text{MC} \exp\left[
\log W(k\Delta)
\mathbf{I}(k\Delta)  
- W(k\Delta) \Delta
\right]\Big\vert_{\xi = \xi_i^{MC}} }
{\sum_{i=1}^{N_\text{MC}}  \exp\left[
\log W(k\Delta)
\mathbf{I}(k\Delta)  
- W(k\Delta) \Delta
\right] \Big\vert_{\xi = \xi_i^{MC}}}.
\eeq

The variance is computed using Gaussian approximation,
\beqs
\sigma^2_{k\vert k} = - \left[ \frac{\partial^2 \log p (\xi_k \vert s_{0,k})}{\partial \xi_k^2} \right]^{-1}
\Big\vert_{\xi_{k\vert k}},
\eeqs \\
{\RV{which after we plug in Eq.~\ref{eq:em:noise_posterior} for the posterior distribution of the noise $p(\xi_k \vert s_{0,k})$ and Eq.~\ref{eq:si:w_ggd_single_ising} for the transition rate $W(k\Delta)$, becomes
\beqs
\sigma^2_{k\vert k}   = 
 \left \lbrace ( \sigma^2_{k \vert k-1} ) ^{-1} + 
\frac{ 4 s^2  e^{2hs}}{ (1+e^{2hs})^2 }
\left[ \mathbf{1}(k\Delta) 
-  \mu  \frac{1- e^{2 hs}}{{1 + e^{2hs} }}
\Delta \right]
\right\rbrace^{-1}.
\eeqs
Alternatively, one can also compute the variance using importance Monte Carlo sampling, similar to how one compute the mean in Eq.~\ref{eq:em:mean_noise_mc}.
}}

\textit{ii. Backwards Kalman smoother}

The estimation for the mean and the covariance of the hidden noise $\xi_k$ can be improved by using the entire observed sequence $s_{0,K}$. Because the hidden noise $\xi$ is generated by a gaussian Markov process, we follow~\cite{smith_brown_2003} and use the fixed interval smoothing algorithm to iteratively update the expected moments at time point $k$ using the expectation value at time point $k+1$:
\beq
\xi_{k\vert K} = \xi_{k \vert k}
+ J_k (\xi_{k+1\vert K} - \xi_{k+1 \vert k} )
\eeq
where 
\beq\label{eq:em:jk}
J_k = a \sigma_{k\vert k}^2 (\sigma_{k+1\vert k}^2)^{-1}
\eeq
and 
\beqs
\sigma^2_{k\vert K} = \sigma^2_{k \vert k}
+ J_k^2 (\sigma^2_{k+1\vert K} - \sigma^2_{k+1 \vert k} ).
\eeqs
The initial condition is $\xi_{k\vert K} = \xi_{k\vert k}$ at the end of the time series, $k = K$. \\

\textit{iii. Compute correlation}

The {\RV{cross-}}correlation of the hidden variable $\xi$ can be computed following the state-space covariance algorithm~\cite{dejong_mackinnon}, which gives us {\RV{the one-step separated correlation}}
\beq
\sigma_{k, k+1 \vert K} = J_k \sigma_{k+1\vert K}^2.
\eeq 
{\RV{The correlation coefficient is
\beqs
\rho_{k, k+1\vert K} = \frac{\sigma_{k, k+1 \vert K}}{( \sigma^2_{k\vert K} )^{1/2} 
( \sigma^2_{k+1\vert K} )^{1/2}}.
\eeqs
}}

{\RV{\subsection{Maximizing data likelihood in the M-step}}}

{\RV{As described in previous paragraphs, in the M(aximization)-step, the complete log likelihood $Q(\bm \theta \vert \bm \theta^{(l)})$ is maximized, and the argument is used as the updated parameter $\bm \theta^{(l+1)}$. }}
{\RV{Different methods are used to}} compute the {\RV{two terms of the}} expected value for the complete log likelihood $Q$, averaged over the posterior distribution of the noise. The terms related to the Gaussian Markov process of the hidden noise is evaluated exactly using the mean and the covariance.  {\RV{For the term contributed by the Glauber dynamics (the first term in Eq.~\ref{eq:si:q}), we approximate it using a second-order Taylor expansion, 
\beqs
Q_g(\bm \theta \vert \bm \theta^{(l)})   = \mathbb{E}_{s_{0,K},\bm{\theta}^{(l)}}[f(\xi_{0, K})] 
 \approx 
 \sum_k f(\xi_{k\vert K}) 
 + \frac{1}{2} \sigma^2_{k\vert K}  \frac{\partial^2 f}{\partial \xi^2}\Big\vert_{\xi = \xi_{k\vert K}},
\eeqs
where for simplicity of notation, we define
\beqs
f(\xi_{0, K}) \equiv \log p(s_{0,K} \vert \xi_{0,K}, \mu).
\eeqs
}}

{\RV{\subsection{Detailed implementation}

In the main text, we applied the EM algorithm to infer the toy model with a single Ising spin with a single exponential decay memory function. Here we give the detailed protocol. To estimate the mean of the posterior distribution of the noise, we use a total number of $N_{MC} = 100$ points in important Monte Carlo sampling for each time step $k$. To compute $Q$, we used 50 random samples of the Gaussian distribution of noise $\xi_{0,K}$, given the mean and the covariance.
}}

{\RV{\subsection{Inferring a GGD model: EM vs heuristic method}

We note that the EM algorithm is not exact, due to logistic form of the Glauber transitioning dynamics, the use of the Gaussian approximation in estimating the posterior distribution of the noise, and the use of Monte Carlo sampling of the posterior noise distribution, which may cause a distortion of the optimization landscape, and leads to problems in convergence (see Chapter 3.4 in~\cite{yuan_thesis}). Furthermore, specifically to our problem where the fluctuation-dissipation relation constrains the parameters, the posterior noise $Q$ is a non-convex function of the parameters to be estimated, $(\mu, a, \sigma^2)$, which can make it hard for the EM algorithm to learn the ground truth parameters, as the EM algorithm approaches sets of parameters where the gradient disappears and not necessarily the global optimum. 

For a single Ising spin, EM recovers the parameters much better than the heuristic method (\sitable ~and \sifigemvsks ~top panels). However, the problem of convergence becomes more severe for EM applied to Potts spins with more than 2 states. The convergence is much slower, and EM does not find the ground truth (see~\sifigemvsks(b) for a sample evolution of EM inferred parameters, for which the existence of the parameters implies existence of multiple EM convergent points). To start from informative parameters, we use the optimal parameters identified by the heuristic method as the initial parameters for the EM algorithm. The EM-identified parameters are then compared to the heuristically identified parameters (see~\sifigemvsks(a) bottom panels). 
}}

\begin{figure*}
\centering
\includegraphics[scale=1]{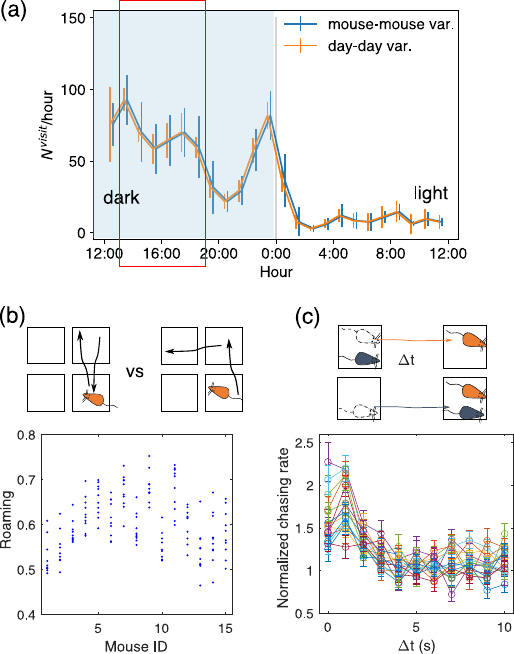}
\caption{Features of non-Markovian dynamics in Eco-HAB mice data. (a) The mean activity varies throughout the day. For our analysis, we focus on the 6 hours of stable activity, indicated by the red rectangle. (b) The exploration measures how likely the mouse keeps its direction of motion across two immediate transitions, and is consistent across days. Specifically, we define \textit{roming} by the ratio of number of consecutive forward transitions (e.g. from box $A$ to $B$ then to a different box $C$) to the number of forward-backward transitions (e.g. from box $A$ to box $B$ and back to box $A$ (c) Mice actively chase each other on a time scale of 5 seconds, here the schematics showing the \textit{gray} mouse chasing the \textit{orange}.   Each colored curve represents one mouse, and the error bars are the standard error from the mean over days. }\label{fig:si:mice_explore_lead_follow}
\end{figure*}

\begin{figure*}
\centering
\includegraphics[scale=1]{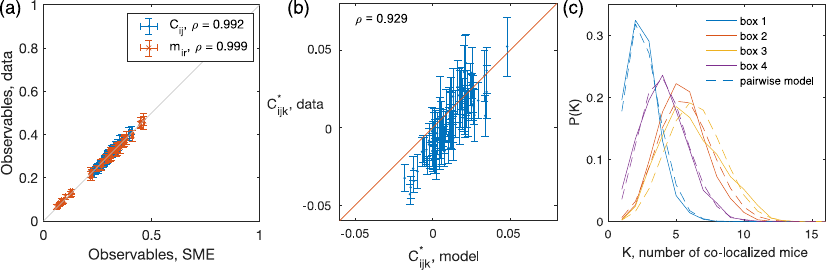}
\caption{Pairwise maximum entropy describes the statics of mice behavior, but not the dynamics. (a) Pairwise maximum entropy model fits the constraints, the mean probability of mouse $i$ in box $r$, and probability that two mice are in the same box $C_{ij}$, and is able to predict higher order structures of the data, here showing the probability of triads of mice in the same box (panel (b)) and the probability of $K$ mice in the same box (panel (c)). Errorbars in panel A and B are the estimation of variablity of the data, by random bootstrapping half of the data. {\RV{The Pearson correlation coefficients $\rho$ are given in the plot.}}  }\label{fig:si:mice_sme}
\end{figure*}

\begin{figure*}
\centering
\includegraphics[width=0.8\textwidth]{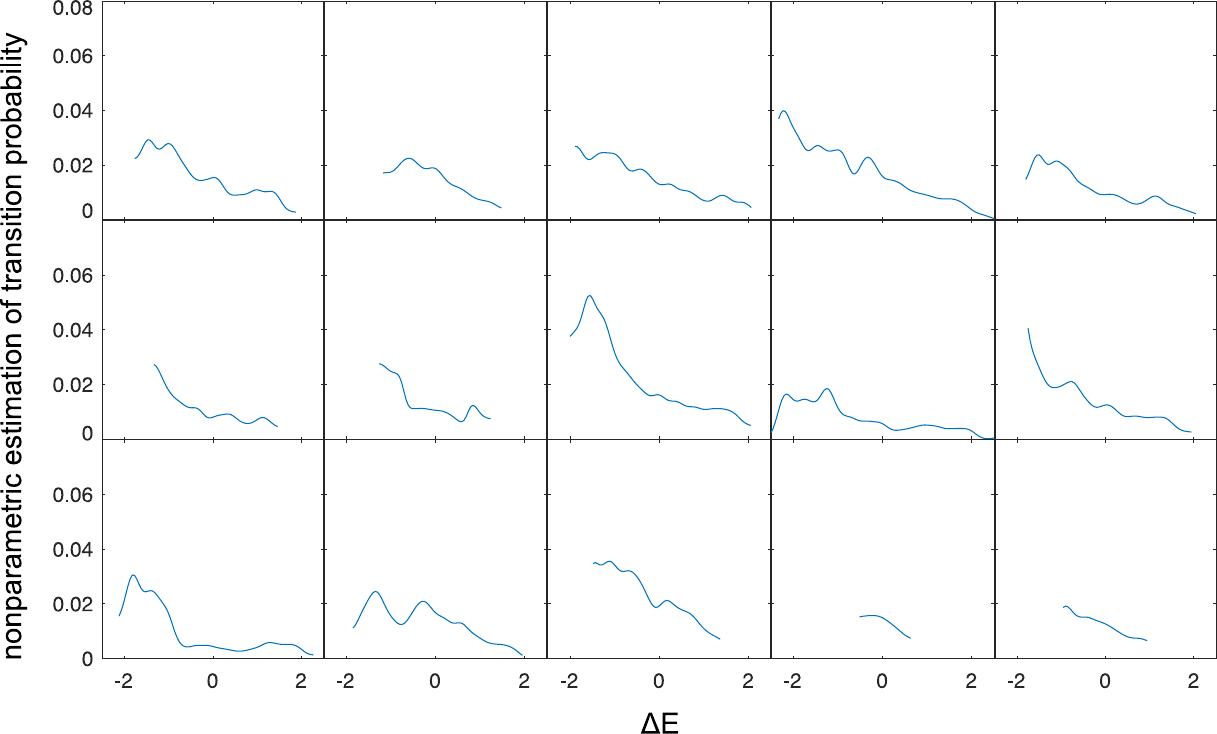}
\caption{{\RV{Nonparametric estimation of the transition probability of each mouse, using Naradaya-Watson kernel estimation with a Gaussian kernel (bandwidth 0.2). }}}
\label{fig:si:nonparam_transprob}
\end{figure*}

\begin{figure*}
\centering
\includegraphics[scale=1]{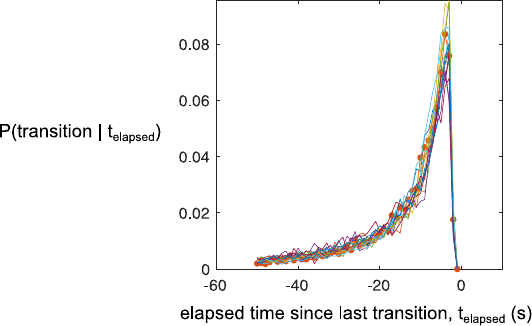}
\caption{The probability of mice transitioning decreases as the time since the last transition increases.}\label{fig:si:mice_eta}
\end{figure*}

\begin{figure*}
\centering
\includegraphics[scale=1]{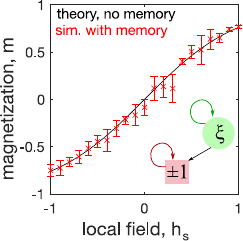}
\caption{The steady state distribution for a single Ising spin with GGD dynamics with exponentailly decaying memory kernel is identical to the Boltzmann distribution, shown here by the magnetization $m = \tanh h$.}\label{si:fig:single_ising_mag}
\end{figure*}

\begin{figure*}
\centering
\includegraphics[width=0.6\textwidth]{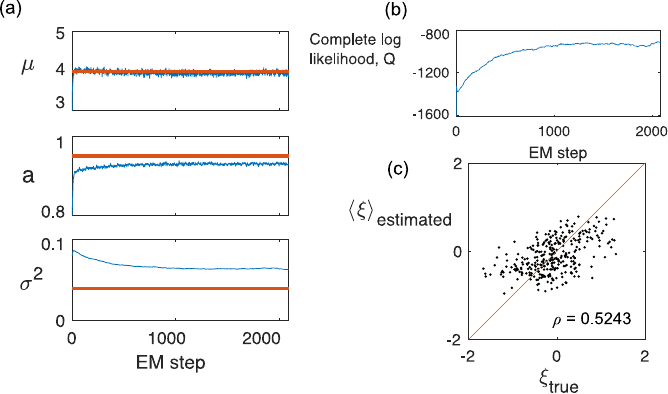}
\caption{The EM algorithm is able to recover the true underlying trajectories{\RV{, shown here for the same simulated Ising spin dynamics as given in Fig.~\ref{fig:ggd_em}}}  (a) The evolution of the parameters as a function of the EM steps. The \textit{red} lines represent the true underlying parameters, and the \textit{blue} curves show the evolution of the inferred parameters. (b) The complete log likelihood of the data and the noise, $Q$, increases throughout the EM iterations. (c) The EM algorithm is able to recover the hidden noise, here plotting the mean of the estimated noise, $\langle\xi\rangle_\text{estimated}$ vs. the true underlying noise $\xi_\text{true}$. {\RV{The Pearson correlation coefficient $\rho = 0.5243$.}}}\label{si:fig:ggd_em_params}
\end{figure*}

\begin{figure*}
\centering
\includegraphics[width=0.6\textwidth]{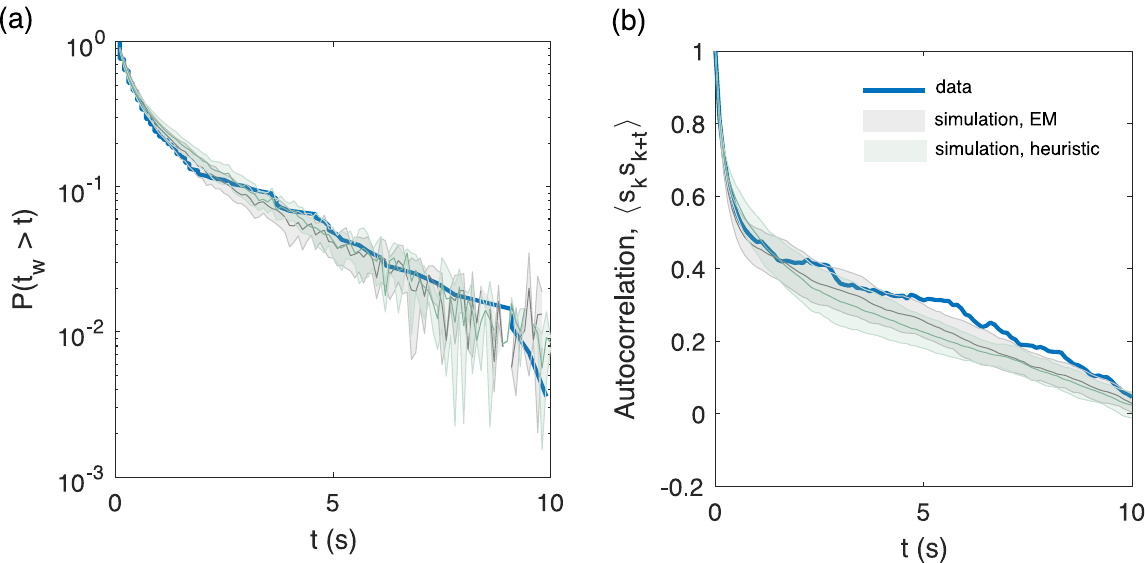}
\caption{Compare GGD inference using the EM algorithm and using the heuristic algorithm minimizing the distance between the waiting time distribution of the data and of simulations. A single Ising spin coupled with a single exponentially-decaying memory kernel is used to generate the data. Simulated trajectories with parameters inferred by the EM algorithm and by the heuristic algorithm recovers both the waiting time distribution (a) and the autocorrelation decay (b) as the data. The envelop{\RV{e}} of the curve is the standard deviation. The results from the EM algorithm is plotted in \textit{gray}, and the results from the heuristic algorithm is plotted in \textit{green}.}\label{si:fig:tw_autocorr_em_ks}
\end{figure*}

\begin{center}
\begin{table}
\centering
\begin{tabular}{c | c  c  c}
& $\mu$ & $a$ & $\sigma^2_\varepsilon$ \\ \hline\hline
true & 4 & 0.95 & 0.04 \\
EM & 3.95 & 0.93 & 0.063 \\
heuristic & 5.75 & 0.86 & 0.28 \\
\end{tabular}
\caption{True vs. inferred parameters for an example GGD dynamics with a single Ising spin and a single exponentially-decaying memory kernel. The two inference methods include the expectation-maximization algorithm, and the heuristic method minimizing the distance of the cumulative distribution of the waiting time between the data and the model.}~\label{si:table:ggd_em}
\end{table}
\end{center}

\begin{figure*}
\centering
\includegraphics[scale=1]{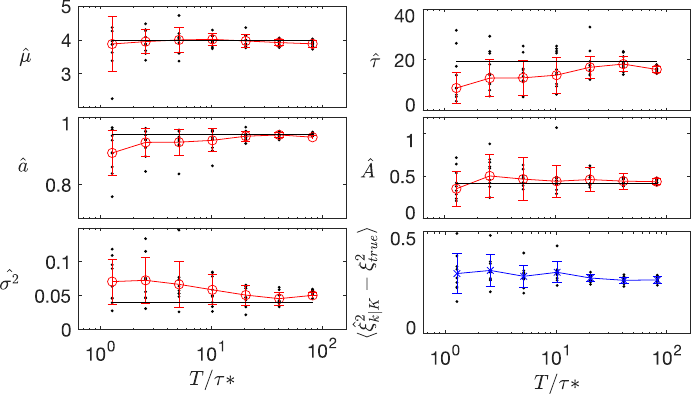}
\caption{{\RV{Accuracy of the EM algorithm as a function of data length $T$, tested for single Ising spin coupled with a single exponentially-decaying memory kernel. The ground truth parameters are given in Table~\ref{si:table:ggd_em}, with $\tau^* = 19.5$ as the ground truth memory timescale. Error bars represent standard deviation across 10 randomly generated sample trajectories for each data length.  (a) The EM inferred parameters approach the ground truth. (b) The inferred memory time scale $\hat{\tau}$ and the strength of memory kernel $\hat{A}$. (c) The mean squared difference between the true noise $\xi$ and the estimated expectation value of the noise $\xi_{k\vert K}$, averaged over the entire trajectory. }}}\label{si:fig:em_error_vs_duration}
\end{figure*}

\begin{figure*}
\centering
\includegraphics[scale=1]{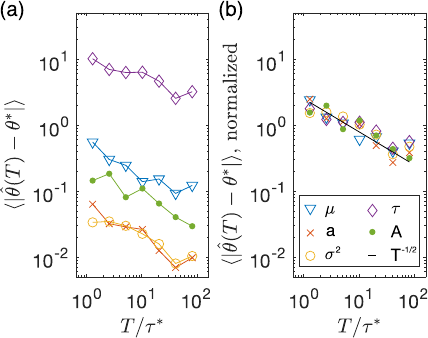}
\caption{{\RV{The error of the EM algorithm scales as $T^{-1/2}$ with respect to the data length $T$. Results shown here are generated from the GGD-EM algorithm for single Ising spins coupled to a single exponentially-decaying memory kernel. (a) The error is measured as the absolute difference between the inferred parameters and the true parameters, averaged over 10 random realizations of the spin trajectories. (b) For the three EM-inferred parameters, $\lbrace \mu, a, \sigma^2 \rbrace$ and the deduced physical parameters $\lbrace \tau, A \rbrace$, the error normalized by its mean over the plotted data duration follows the same scaling relation. }}}\label{si:fig:em_error_scaling}
\end{figure*}

\begin{figure*}
\centering
\includegraphics[scale=1]{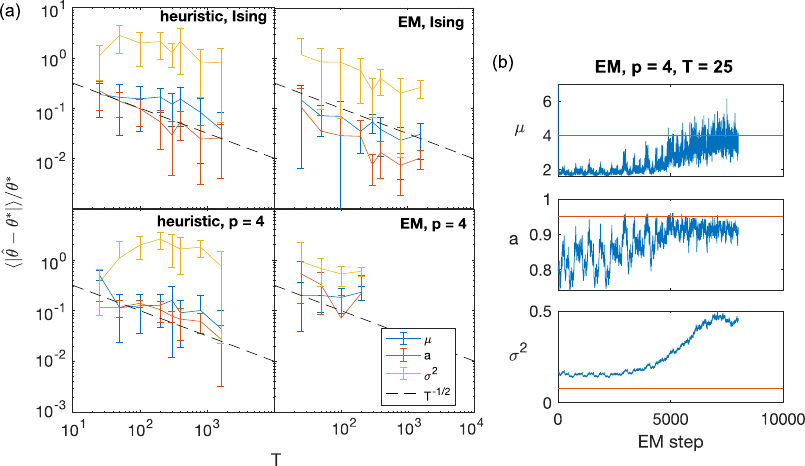}
\caption{{\RV{\textbf{Performance comparison of the EM algorith and the heuristic method.} (a) Inference error of the EM algorithm and the heuristic method plotted against data length $T$, for a single Ising spin with a single memory decay ($\mu* = $,$a* = 0.95$, $\sigma^2_* =  0.04$), and for a single Potts spin ($p = 4$) with an exponentially-decaying memory kernel (with groundtruth $\mu* = 4$, $a* = 0.95$, $\sigma^2_* =  0.08$). Error bars represent standard deviation across 10 randomly generated sample trajectories for each data length.  The intial parameters of the EM alogrithm is set to the parameters identified using the heuristic algorithm for each trial. (b) The evolution of parameters as a function of the EM steps for an example trial for a single Potts spin ($p = 4$) with data length $T = 25$. The two plateaus imply multiple convergence points for the EM algorithm when applied to single Potts with $p = 4$ states.   }}}\label{si:fig:em_vs_heuristic_ising_potts4}
\end{figure*}

\begin{figure*}
\centering
\includegraphics[scale=1]{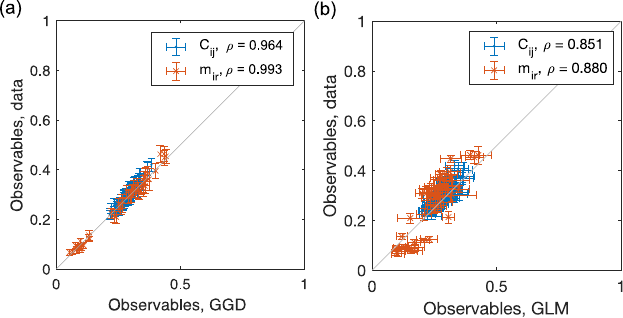}
\caption{By construction, the GGD model atop of pairwise maximum entropy model reproduces the mean and the pairwise correlation {\RV{(panel (a)), while GLM does not (panel (b)).}}  Error bars are the bootstrapped data variation.}\label{si:fig:ggd_mice_static}
\end{figure*}

\begin{figure*}
\centering
\includegraphics[scale=1]{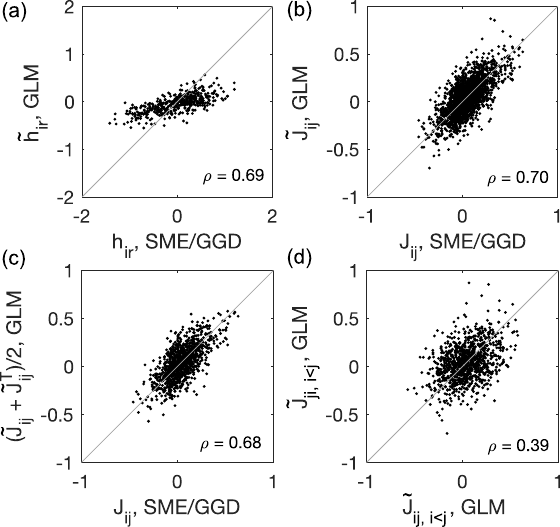}
\caption{{\RV{Comparison of parameters inferred using the pairwise maximum entropy model (SME /GGD) and parameters inferred using the generalized linear model (GLM) which does not impose detailed balance. Inferred parameters plotted include the local field (panel (a)), the pairwise interaction (panel(b)). The symmetrized GLM interaction correlates with the SME/GGD interactions (panel (c)). The asymmetricity of the inferred GLM interaction is shown by comparing the $\widetilde{J}_{ij}$ and $\widetilde{J}_{ji}$ for each $i < j$ pair (panel (d)). The Pearson correlation coefficients are given in the plot.   }}}
\label{fig:si:glm_vs_sme_params}
\end{figure*}

\begin{figure*}
\centering
\includegraphics[scale=1]{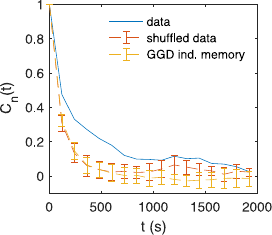}
\caption{The GGD model with independent memory is unable to reproduce the emergent long time scale in the correlation function of occupation number, i.e. number of mice in the same box, suggesting there is memory dependence on other mice.}\label{si:fig:ggd_mice_autocorr}
\end{figure*}

\end{document}